\documentclass[12pt]{article}
\usepackage{amssymb}
\usepackage{amsmath}
\usepackage{epsfig}
\usepackage{color}

\newcommand{\bk}{{(k)}}

\newcommand{\D}{{\mathcal D}}
\renewcommand{\Re}{Re}
\renewcommand{\Im}{Im}
\newcommand{\fer}[1]{(\ref{#1})}

\renewcommand{\H}{{\mathcal H}}
\renewcommand{\S}{{\mathcal S}}
\renewcommand{\O}{{\mathcal O}}
\newcommand{\B}{{\mathcal B}}
\renewcommand{\L}{{\mathcal L}}

\renewcommand{\a}{\alpha}
\newcommand{\R}{{\mathcal R}}
\renewcommand{\S}{\Sigma}

\newcommand{\h}{{\mathbf h}}
\newcommand{\F}{{\mathcal F}}

\newcommand{\unit}{{\mathbf 1}}

\newcommand{\M}{{\mathcal M}}
\newcommand{\N}{{\mathcal N}}

\newcommand{\C}{{\mathcal C}}

\newcommand{\unitD}{{\overline{D}}}
\newcommand{\unitOmega}{{\overline{\Omega}}}

\newcommand{\PV}{{\mathcal PV}}

\newcommand{\sfP}{{\sf P}}

\begin{document}
\title{Cyclic thermodynamic processes \\ and entropy production}
\author{
Walid K. Abou Salem and J\"urg Fr\"ohlich \\
Institut f\"ur Theoretische Physik, ETH-H\"onggerberg \\
CH-8093 Z\"urich, Switzerland \footnote{\it e-mail: walid@itp.phys.ethz.ch, juerg@itp.phys.ethz.ch}}
\date{\; }

\maketitle

\begin{abstract}
We study the time evolution of a periodically driven
quantum-mechanical system coupled to several reservoirs of free
fermions at different temperatures. This
is a paradigm of a cyclic thermodynamic process. We introduce the
notion of a Floquet Liouvillean as the generator of the dynamics
of the coupled system on an extended Hilbert space. We show that
the time-periodic state which the state of the coupled system
converges to after very many periods corresponds to a zero-energy
resonance of the Floquet Liouvillean. We then show that the
entropy production per cycle is (strictly) positive, a property that implies
Carnot's formulation of the second law of thermodynamics.
\end{abstract}

\setcounter{footnote}{1}


\section{Introduction}
During the past several years, there has been substantial progress
in the program of deriving the fundamental laws of thermodynamics
from nonequilibrium quantum statistical mechanics (see [A-SF1, A-S] for
a synopsis). In this paper, we make a contribution to this program
by studying Carnot's formulation of the second law of
thermodynamics from the point of view of quantum statistical
mechanics. For the sake of concreteness, we consider a
periodically driven two-level quantum-mechanical system, $\S,$
coupled to $n\ge 2$ reservoirs of free fermions, $\R_1, \cdots
,\R_n.$ Our analysis can be generalized to a system composed of an
arbitrary quantum-mechanical system with a finite-dimensional
Hilbert space coupled to several reservoirs of (free) bosons or
fermions at different temperatures.\footnote[1]{Bosonic reservoirs are more difficult,
technically, since the interaction term coupling $\S$ to the bosonic
reservoirs is generally an unbounded operator. However, one may readily extend the methods developed in [MMS1,2] to study the bosonic case.} In
order to study the time evolution of the coupled system, we extend
Floquet theory for periodically driven quantum systems at zero
temperature (see for example [Ho, Ya1, Ya2]) to apply to systems
at positive temperatures.

In particular, we introduce an operator, the Floquet Liouvillean,
generating the dynamics of the system on an extended Hilbert
space, and we show that the time-periodic state, which the 
state of the system converges to after very many periods,
corresponds to a zero-energy resonance of the Floquet Liouvillean.
We also prove (strict) positivity of entropy production per cycle,
which amounts to Carnot's formulation of the second law of
thermodynamics. For weak enough coupling of the small system,
$\S,$ to the reservoirs, the time-periodic state corresponding to
the zero-energy resonance can be expanded in powers of the
coupling constant. This is of considerable practical importance,
since it enables one to explicitly compute the {\it degree of
efficiency} of the coupled system. Further discussion of the
second law of thermodynamics and another proof of convergence to
time-periodic states using methods of scattering theory will
appear in [A-SF1]. As far as we know, all previous
investigations of Carnot's formulation of the second law of
thermodynamics from the point of view of quantum statistical
mechanics assumed that the coupling is switched on at time
$t=t_0,$ and then switched off at a later time $t=t_0+\tau$ (see
[PW, BR] and references therein). The novelty of our approach is
to prove that the state of the coupled system converges to a
time-periodic state with the same period as the one of the
interaction, and that entropy production is (strictly) positive; (see also [FMSUe,MMS1,2]).

The organization of this paper is as follows. In section 2, we
recall some basic notions from quantum statistical mechanics, in
particular, time-dependent perturbations of $C^*$-dynamical
systems, relative entropy, and Carnot's formulation of the second
law of thermodynamics. We also discuss sufficient conditions to
prove strict positivity of entropy production per cycle. These
conditions are satisfied in the concrete example we consider in
the following section. In section 3, we discuss a concrete model, and
we state the assumptions we make on the interaction between the
small system, $\S,$ and the reservoirs, $\R_1,\cdots ,\R_n.$ In
section 4, we introduce the Floquet Liouvillean, whose spectrum
we study using complex spectral deformation techniques.
In section 5, we use results on the Floquet
Liouvillean to prove convergence of the 
state of the coupled system to a time-periodic state. This is one
of the main results of our paper. We then prove strict positivity
of entropy production per cycle in section 6, and we discuss how to
compute the degree of efficiency of the coupled system for weak enough
coupling. The main ingredients of our analysis are a concrete
representation of the fermionic reservoirs (Araki-Wyss
representation), a spectral approach to cyclic thermodynamic
processes using the so called Floquet Liouvillean, and complex spectral
deformation techniques.

\bigskip
\noindent {\bf Acknowledgements}
\noindent W.A.S. would like to thank Gian Michele Graf for enjoyable discussions.


\section{General considerations}

In this section, we recall some of the basic notions of quantum
statistical mechanics, and we introduce the notion of
time-periodic states. Although some of the material is standard, it
is presented in this section in order to make our exposition reasonably self-contained.

In the algebraic formulation of quantum statistical mechanics, a
physical system is described by a $C^*$ or $W^*-$ dynamical
system. Since we only consider fermionic reservoirs in this paper,
we restrict our attention to the discussion of $C^*-$dynamical
systems. However, our analysis can be generalized to
$W^*-$dynamical systems; (see for example [DJP], and also [MMS1,2]).

A $C^*$-dynamical system is a pair $(\O,\a),$ where $\O$ is the
kinematical algebra of the system, a $C^*$-algebra with
identity, and $\a,$ which specifies the time evolution, is a
norm-continuous one-parameter group of $*$-automorphisms of $\O.$ A physical
state of the system is described by a positive, linear functional
$\omega,$ with $\omega(\unit)=1.$ The set $E(\O)$ of all
states is a convex, weak-* compact subset of the dual $\O^*.$

{\it Physically relevant} states of (isolated) thermal reservoirs are assumed to be
normal to equilibrium states characterized by the Kubo-Martin-Schwinger (KMS)
condition. An equilibrium state of $(\O,\a)$ at inverse
temperature $\beta, \omega_\beta,$ is an
$(\a,\beta)$-KMS state satisfying
$$\omega_\beta (a\a^t(b))=\omega_\beta(\a^{t-i\beta}(b)a)\; ,$$
for all $a,b\in\O^0,$ where $\O^0$ is norm-dense in $\O.$

We briefly recall the perturbation theory for $C^*-$dynamical systems; (for
further details, see for example [BR]). Let $\delta$ be the
generator of $\a,$ ie, $\a^t=e^{t\delta}, t\in {\mathbf R}.$ The
domain of the derivation $\delta, \; \D(\delta),$ is a $*$-subalgebra of $\O,$
and for all $a,b \in \D(\delta),$
\begin{equation*}
\delta(a)^* = \delta(a^*) , \; \delta (ab)= \delta (a)b + a\delta (b) \; .
\end{equation*}

Consider the time-dependent family of perturbations, $\{ gV(t)
\}_{t\in {\mathbf R}}$, with selfadjoint elements $gV(t)\in\O.$
Then $\a_g ,$ the perturbed time evolution, is a norm-continuous one-parameter family of
$*$-automorphisms of $\O$ satisfying
\begin{equation}
\frac{d}{dt}\a_g^t (a) = \a_g^t (\delta (a) + ig [V(t), a]) \; ,
\end{equation}
and $\a_g^0 (a)= a ,$ for all $a\in \O.$ Explicitly,

\begin{equation}
\a_g^t (a) = \a^t (a) + \sum_{n\ge 1} i^n g^n \int_{0}^t dt_1 \int_{0}^{t_1}dt_2 \cdots \int_{0}^{t_{n-1}}dt_n  [\a^{t_n}(V(t_n)), \cdots , [\a^{t_1}(V(t_1)), \a^t(a)], \cdots ] .
\label{perturbedevolution}
\end{equation}

In the standard interaction picture,
$$\a_g^t (a)= \Gamma_g^t \a^t (a)\Gamma_g^{t*},$$
where $\Gamma_g$ is a unitary element of $\O$ which satisfies
\begin{equation}
\label{intevol}
\frac{d}{dt}\Gamma_g^t=i\Gamma_g^t \a^t (gV(t)),
\end{equation}
and $\Gamma_g^0=1 \; ;$ ie,
$$\Gamma_g^t = \unit + \sum_{n\ge 1} i^n g^n \int_0^t dt_1 \int_0^{t_1}dt_2 \cdots \int_0^{t_{n-1}}dt_n \a^{t_n}(V(t_n))\cdots \a^{t_1}(V(t_1)) .$$

Next, we discuss the notion of relative entropy and of entropy
production. Assume that there exists a reference $C^*$-dynamics
$\sigma_{\omega}$ on $\O$ and a state $\omega$ with the property that $\omega$ is an
$(\sigma_\omega, -1)$-KMS state. (Equivalently, at inverse temperature $\beta\ge 0,
\sigma_{\omega, \beta}^t=\sigma_\omega^{-t/\beta}$). Let
$\delta_\omega$ be the generator of $\sigma_\omega,$ and let
$(\H_\omega, \pi_\omega , \Omega_\omega)$ be the GNS
representation of the kinematical algebra $\O$ associated to the
state $\omega .$ (For further discussion of the GNS
construction see, for example, [BR].)

A state $\eta\in E(\O)$ is called $\omega$-normal if there exists
a density matrix $\rho_\eta$ on $\H_\omega,$ such that, for all
$a\in \O,$
$$\eta(a)=Tr(\rho_\eta \pi_\omega (a)),$$
where $Tr$ is the trace over $\H_\omega.$
We will denote by $\N_\omega$ the set of all $\omega$-normal
states in $E(\O).$

For a state $\eta\in \N_\omega,$ which might be {\it time-dependent}, denote by $Ent(\eta | \omega)$ the
relative entropy of Araki [Ar,D]. For finite systems,
\begin{equation}
Ent(\eta | \omega )= -Tr(\eta \log \omega -\eta \log\eta).
\end{equation}
If $\eta\not\in \N_\omega,$ set $Ent(\eta |\omega)=+\infty .$\footnote{Note the choice of the sign of relative entropy.}

For a self-adoint perturbation $gV(t)\in \D(\delta_\omega)$ of the
dynamical system, as discussed above, we define the rate of entropy
production in a state $\eta\in E(\O)$ relative to a reference
state $\omega$ as
\begin{equation}
\label{entprod}
Ep(\eta):= \eta (\delta_\omega (g V(t))) ;
\end{equation}
see for example [BR,JP4]. 


It is instructive to see how one obtains this expression for
entropy production as the thermodynamic limit of quantities referring to
 finitely extended reservoirs.[A-SF2]
Consider a quantum system composed of a small system, $\S,$
coupled to $n$ reservoirs, $\R_1,\cdots ,\R_n,$ at inverse
temperatures $\beta_1,\cdots, \beta_n,$ respectively. We first treat the
reservoirs as finitely extended systems and then take the thermodynamic limit of
suitable quantities. The total Hamiltonian of the {\it finite}
coupled system is
$$H(t)= H^\S + \sum_{i=1}^n H^{\R_i} + gV(t) ,$$
where $H^\S$ is the Hamiltonian of the uncoupled small system,
$H^{\R_i}$ is the Hamiltonian of the $i^{th}$ uncoupled reservoir,
and $gV(t)$ is the interaction term coupling $\S$ to the reservoirs. Let
the reference state of the reservoirs be $\omega^{\R}$ with corresponding density matrix $\rho^\R$
given by $$\rho^\R=\rho^{\R_1}\otimes \cdots\otimes \rho^{\R_n},$$
where $\rho^{\R_i}$ is the density matrix corresponding to the equilibrium state
of the reservoir $\R_i$ at inverse temperature $\beta_i$, which is given by
$$\rho^{\R_i}=\frac{e^{-\beta_iH^{\R_i}}}{Tr_{\R_i}(e^{-\beta_iH^{\R_i}})},$$
where $Tr_{\R_i}$ is the trace over the Hilbert space of the reservoir $\R_i.$

Define $\omega_t := \omega\circ \a_g^t,$ where $\omega$ is the initial $(\sigma_\omega,-1)$-KMS state, and let
$\rho^\omega$ be the density matrix corresponding to $\omega.$
For a finite system,
\begin{align*}
Ent (\omega_t | \omega^\R) &= -Tr (\rho^{\omega_t}\log \rho^\R)+Tr(\rho^{\omega_t} \log\rho^{\omega_t}) \\
&= -Tr (\rho^{\omega_t}\log\rho^{\R})+Tr(\rho^{\omega}\log\rho^{\omega}) ,
\end{align*}
and hence
$$\frac{d}{dt}Ent(\omega_t|\omega^\R )=i Tr ([H(t), \rho^{\omega_t}]\log\rho^{\R}) .$$

By cyclicity of the trace we have that

\begin{align*}
\frac{d}{dt}Ent(\omega_t|\omega^\R )
&=i\sum_{i=1}^n\beta_i Tr (\rho^{\omega_t} [H(t), H^{\R_i}]) \\
&=i \sum_{i=1}^n \beta_i Tr (\rho^{\omega_t} [gV(t), H^{\R_i}])\\
&=\omega \circ\a_g^t (\delta_\omega (gV(t))),
\end{align*}
where we have used in the last equation that $\delta_\omega=-\sum_{i=1}^n\beta_i\delta_i,$ and $\delta_i=i[H^{\R_i},\cdot].$
Note that the thermodynamic limit of the entropy production rate is
well-defined.

One may relate the entropy production to the heat flux from the
reservoirs. The heat flux from reservoir $\R_i, i=1,
\cdots ,n,$ at time $t$ is
\begin{eqnarray}
\Phi^{\R_i}(t)
&:=& -\frac{d}{dt}\omega \circ \a_g^t (H^{\R_i}) \\
&=& -i\omega (\a_g^t ([gV(t), H^{\R_i}]))\nonumber \\
&=& \omega \circ \a_g^t (\delta_i (gV(t))) .
\end{eqnarray}

It follows that
\begin{equation}
\sum_{i=1}^n \beta_i \Phi^{\R_i}(t) = -Ep(\omega\circ \a_g^t) ,
\label{2ndlaw}
\end{equation}
which, for $n=1$  and reversible processes, is a familiar equation.

We now give a definition of a time-periodic state in case of a time-periodic coupling.

\bigskip
\noindent {\bf Time-periodic state.}\footnote{A weaker definition
of a time-periodic state is
$$\omega_{g,s}^+ :=\lim_{n\rightarrow\infty}\frac{1}{n\tau}\int_0^{n\tau}dt \omega\circ \a_g^{t+s}.$$}
\noindent{\it Assume that the perturbation $gV(t)$ is time periodic with period
$\tau,$ and is norm-differentiable, for $t>0.$ For $s\in [0,\tau),$
define the time-periodic state $\omega_{g,s}^+$ as
\begin{equation}
\label{timeper}
\omega_{g,s}^+ := \lim_{n\rightarrow\infty} \omega\circ\a_g^{n\tau +s} .
\end{equation}}
Note that it follows from this definition of a time-periodic state that,
\begin{equation*}
\omega_{g,s}^+\circ \a_g^\tau = \omega_{g,s}^+ .
\end{equation*}
We will show in section 5 that this state is related to a zero-energy resonance of the so called Floquet Liouvillean.

Next, we exhibit a connection to Carnot's formulation of the
second law of thermodynamics. Consider a cyclic thermodynamic
process in which $\S$ is coupled to two reservoirs, $\R_1$ and
$\R_2,$ at temperatures $T_1$ and $T_2,$ respectively, with
$T_1>T_2.$ Reservoir 1 acts as a heat source and reservoir 2 as a
heat sink. 
Recall that the generator of the free dynamics of reservoir $\R_i$ is $\delta_i,i=1,2.$
Since $V(t)$ is norm differentiable, for $t>0,$ it follows that
$$\frac{d}{dt}\delta_i (\Gamma_g^t) = \delta_i (\frac{d}{dt} \Gamma_g^t) ,$$
for $i=1,2.$ Using the fact that $\delta_i, i=1,2,$ is a
$*$-derivation which commutes with $\a^t$ and equation
\fer{intevol}, it follows that
$$\frac{d}{dt}\omega (\Gamma_g^t \delta_i (\Gamma_g^{t*})) = i \omega\circ \a_g^t (\delta_i (gV(t))).$$

Therefore, the heat energy flowing from reservoir $\R_i$ into system $\S$ during the time interval $[0,t]$ is
\begin{align}
\Delta_0^t Q_i &= -\int_0^t dt' \omega\circ\a_g^{t'} (\delta_i (g V(t'))) \\
&= -i \omega(\Gamma_g^t \delta_i (\Gamma_g^{t*})) \;, \label{heat}
\end{align}
$i=1,2.$
Moreover, integrating \fer{2ndlaw}, we have
\begin{equation}
\label{carnot1}
\beta_1 \Delta_0^t Q_1 + \beta_2 \Delta_0^t Q_2=-Ent (\omega\circ \a_g^t |\omega) \le 0,
\end{equation}
since the relative entropy $Ent(\omega\circ\a_g^t |\omega)\ge
0.$\footnote{This follows from a general trace inequality, see
for example [BR].}


Now define the heat flow per cycle from each reservoir $\R_i$ into $\S$ as
\begin{equation}
\Delta Q_i := \lim_{n\rightarrow\infty} [Q_i((n+1)\tau)-Q_i(n\tau) ] , i=1,2.
\end{equation}

We assume that the system converges to a time-periodic state, and that during every cycle, it performs work, ie,
\begin{equation}
\label{work}
\Delta A =\Delta Q_1 + \Delta Q_2 \ge 0.
\end{equation}
It follows from the definite sign of relative entropy and the existence of the time-periodic limit that the entropy production per cycle is nonnegative,
\begin{equation}
\label{EntropySign}
\Delta Ent =\int_0^\tau dt \omega_{g,t}^+ (\delta_\omega (gV(t)))=-(\beta_1\Delta Q_1 + \beta_2\Delta Q_2)\ge 0.
\end{equation}
The fact that $\beta_1\le \beta_2,$ \fer{work} and \fer{EntropySign}, imply that $\Delta Q_1\ge 0.$
It then follows that the {\it degree of efficiency},
\begin{align}
\eta &:= \frac{\Delta A}{\Delta Q_1} \\
&=\frac{\Delta Q_1 + \Delta Q_2}{\Delta Q_1} \\
&\le \frac{T_1 - T_2}{T_1}=: \eta^{Carnot},
\end{align}
which is nothing but Carnot's formulation of the second law of
thermodynamics. In certain situations, one can show that
\begin{equation}
\label{EntCycle}
\lim_{n\rightarrow\infty}[Ent(\omega\circ\a_g^{(n+1)\tau}|\omega)-Ent(\omega\circ\a_g^{n\tau}|\omega)]>0,
\end{equation}
which holds for the model we consider in this paper, and hence $\eta<\eta^{Carnot}.$ The following proposition states sufficient conditions for inequality \fer{EntCycle} to hold.

\bigskip
\noindent {\bf Proposition 2.1}

{\it For $t\in {\mathbf R}^+,$ let $s:=t \; mod \; \tau.$ Suppose that
\begin{itemize}
\item[(a)] $\omega_{g,s}^+ \not\in \N_\omega,$ and that
\item[(b)] $\sup_{T\in {\mathbf R}^+}|\int_0^T dt \{\omega_{g,t\; mod \; \tau}^+ (\delta_\omega (gV(t)))-\omega\circ\a_g^t (\delta_\omega (gV(t)))  \}|<C,$
where $C$ is a finite, nonnegative constant.
\end{itemize}
Then $Ep(\omega_{g,s}^+)>0.$ }
\bigskip

\noindent {\it Proof.} Suppose that $Ep(\omega_{g,s}^+)=0.$ Then
\begin{align*}
Ent(\omega\circ\a_g^t|\omega)&= \int_0^t dt' \omega\circ \a_g^{t'} (\delta_\omega (gV(t'))) \\
&= \int_0^t dt' \{ \omega\circ \a_g^{t'} (\delta_\omega (gV(t')))-\omega_{g,t'\; mod \; \tau}^+ (\delta_\omega (gV(t')))\} \\
\le C.
\end{align*}
In particular,
$$Ent(\omega_{g,s}^+|\omega)=\lim_{n\rightarrow\infty} Ent (\omega\circ \a_g^{n\tau +s}|\omega)\le C.$$
Let $\M=\pi_\omega(\O)'',$ the double commutant of $\pi_\omega(\O),$ and let $\M_*$ be its predual. The set of all states $\gamma\in \N_\omega$ such that
$Ent(\gamma|\omega)\le C$ is $\sigma(\M_*,\M)$-compact (see [BR,D]).  It
follows that $\omega_{g,s}^+\in \N_\omega,$ which contradicts
assumption (a). $\Box$

\bigskip

\noindent{\bf Summary of main results}

Before specifying the concrete model we study, we briefly describe the main results of this paper, deferring precise  statements and proofs to subsequent sections. One key result of this paper is transposing the problem of proving convergence to a time-periodic state to a spectral problem by introducing the so called Floquet Liouvillean (Section 4).  In Section 5, Theorem 5.1, we show that the time-periodic state to which the state of the coupled system converges after very many periods is related to a zero-energy resonance of the Floquet Liouvillean. We also establish {\it strict} positivity of entropy production per cycle in the time-periodic state (Section 6, Theorem 6.3). In the case of two reservoirs, positivity of entropy production implies Carnot's formulation of the second law of thermodynamics,
\begin{equation*}
\eta<\eta^{Carnot}.
\end{equation*}
Our analysis also has some quantitative implications: At weak coupling, the time-periodic state is analytic in the coupling constant, and hence one can calculate the entropy production per cycle perturbatively. This leads to a perturbative calculation of the degree of efficiency when the system is operated as a heat engine.


\section{The Model}

As an example, we consider a two-level quantum system $\S$ coupled to $n$ reservoirs $,\R_1,\cdots ,\R_n,
n\ge 2,$ of free fermions in thermal equilibrium at inverse
temperatures $\beta_1, \cdots, \beta_n,$ and chemical potentials
$\mu_1, \cdots, \mu_n.$ \footnote{For the sake of simplicity of exposition, we set the chemical potentials of the reservoirs to be equal in the subsequent sections.}


\subsection{The small system}

The kinematical algebra of $\S$ is $\O^\S=\M({\mathbf C}^2)$, the
algebra of complex $2\times 2$ matrices over the Hilbert space
$\H^\S={\mathbf C}^2$. Its Hamiltonian is given by $H^\S=\omega_0
\sigma_3$, where $\sigma_i,i=1,2,3,$ are the Pauli matrices. When
the system $\S$ is not coupled to the reservoirs, its dynamics in
the Heisenberg picture is given by
\begin{equation}
\a_\S^t (a) := e^{iH^\S t} a e^{-iH^\S t} \; ,
\end{equation}
for $a\in \O^\S.$

A physical state of the small system is described by a density
matrix $\rho_\S.$ The operator
$\kappa_\S=\rho_\S^{1/2}$ belongs to the space of Hilbert-Schmidt
operators, which is isomorphic to $\H^\S\otimes\H^\S.$ Two
commuting representations of $\O^\S$ on $\H^\S\otimes\H^\S$ are
given by
\begin{align}
& \pi_\S (a) := a\otimes\unit^\S \; , \\
& \pi_\S^\# (a) := \unit^\S \otimes C^\S a C^\S \; ,
\end{align}
where $C^\S$ is an antiunitary involution on $\H^\S$
corresponding to complex conjugation; (see for example [BFS]).

The generator of the free dynamics on the Hilbert space
$\H^\S\otimes\H^\S$ is the standard Liouvillean
\begin{equation}
\L^\S = H^\S\otimes \unit^\S - \unit^\S\otimes H^\S \; .
\end{equation}

The spectrum of $\L^\S$ is $\sigma (\L^\S)=\{ -2\omega_0,0,2\omega_0 \}
,$ with double degeneracy at zero.

Let $\omega^\S$ be the initial state of the small system $\S,$
with corresponding vector $\Omega^\S\in \H^\S\otimes\H^\S.$ The
modular operator associated with $\omega^\S$ is $\Delta^\S =
\omega^\S\otimes \overline{\omega^\S}^{-1},$ and the modular
conjugation operator, $J^\S,$ is given by
$$J^\S (\phi\otimes\psi)=\overline{\psi}\otimes \overline{\phi},$$
for $\phi,\psi\in \H^\S.$
If $\omega_\S$ corresponds to the trace state, then $\Delta^\S=\unit^\S\otimes\unit^\S.$


\subsection{The reservoirs}

Each thermal reservoir is formed of free fermions. It is
infinitely extended and {\it dispersive}. We assume that the
Hilbert space of a single fermion is $\h = L^2 ({\mathbf R}^+,m(u)du ;\B),$
where $\B$ is an auxiliary Hilbert space, and $m(u)du$ is a measure on ${\mathbf R}^+$.
We also assume that the single-fermion Hamiltonian, $h,$ corresponds to the operator of
multiplication by $u\in {\mathbf R}^+ .$ For instance, for
reservoirs formed of nonrelativistic fermions in ${\mathbf R}^3,$
the auxiliary Hilbert space $\B$ is $L^2 (S^2, d\sigma),$ where
$S^2$ is the unit sphere in ${\mathbf R}^3,d\sigma$ is the uniform
measure on $S^2,$ and $u=|\vec{k}|^2,$ where $\vec{k}\in {\mathbf
R}^3$ is the particle's momentum. In the latter case, the measure
on ${\mathbf R}^+$ is choosen to be $m(u)du=\frac{1}{2}\sqrt{u}du.$ For the sake of concreteness, we will consider $\B=L^2(S^{d-1}, d\sigma),d>2,$ in the sequel. 

Let $b$ and $b^*$ be the annihilation-and creation operators on
the Fermionic Fock space $\F(L^2({\mathbf R}^+,\B)).$ They satisfy the CAR
\begin{align}
&\{ b^\# (f), b^\# (g) \} = 0 \; , \\
&\{ b(f),b^* (g) \} = (f,g)\unit \; ,
\end{align}
where $b^\#$ stands for $b$ or $b^*,$ $f,g\in L^2 ({\mathbf R}^+ ;
\B),$ and $(\cdot , \cdot)$ denotes the scalar product in $L^2
({\mathbf R}^+ ; \B).$ Moreover, let $\Omega^\R$ denote the vacuum
state in $\F(L^2({\mathbf R^+};\B)) .$

The kinematical algebra, $\O^{\R_i},$ of the $i^{th}$ reservoir $\R_i
,i=1,\cdots ,n,$ is generated by $b_i^\#$ and the identity
$\unit^{\R_i}.$ The free dynamics of each reservoir (before the
systems are coupled) is given by
\begin{equation}
\a_{\R_i}^t (b_i^\# (f)) = b_i^\# (e^{itu}f) \; ,
\end{equation}
for $i=1,\cdots ,n, f\in L^2 ({\mathbf R}^+ ; \B).$ For a
nonzero chemical potential, $\mu_i,$ of reservoir $\R_i ,$ an auxiliary free
dynamics is generated by $\tilde{H}^{\R_i}=d\Gamma_i
(h-\mu_i)$; see for example [BR].

The ($\a_{\R_i}^t, \beta_i, \mu_i$)-KMS state,
$\omega^{\R_i},$ of each reservoir $\R_i, i=1,\cdots , n,$ at
inverse temperature $\beta_i$ and chemical potential $\mu_i$, is
the gauge invariant, quasi-free state uniquely determined by the
two-point function
\begin{equation}
\label{twopoint}
\omega^{\R_i} (b_i^*(f)b_i(f)) = (f, \rho_{\beta_i,\mu_i }(\cdot ) f) \; ,
\end{equation}
where $\rho_{\beta_i,\mu_i }(u) := \frac{1}{e^{\beta_i (u-\mu_i)}+1}.$

Next, we introduce $\F_i^{AW}:=\F^{\R_i}(L^2({\mathbf R}^+;\B))\otimes \F^{\R_i}(L^2({\mathbf R}^+;\B)),$ the GNS Hilbert space
for the Araki-Wyss representation of each fermionic reservoir $\R_i$ associated with the state $\omega^{\R_i},$ [ArWy].
Let $\tilde{b}_i$ and $\tilde{b}_i^*$ denote the annihilation- and creation operators on $\F^{\R_i}(L^2({\mathbf R}^+;\B))$
satisfying the CAR, and denote by $\Omega^{\R_i}$ the vacuum state in $\F^{\R_i}(L^2({\mathbf R}^+;\B)),$ with
$\tilde{b}_i\Omega^{\R_i}=0.$ The Araki-Wyss representation, $\pi_i,$ of the kinematical algebra $\O^{\R_i},
i=1,\cdots, n,$ on $\F^{AW}_i$  is given by
\begin{eqnarray}
\pi_i (b_i (f)) &:=& \tilde{b}_i (\sqrt{1-\rho_{\beta_i,\mu_i}} \; f) \otimes \unit^{\R_i} + (-1)^{N_i}\otimes \tilde{b}_i^*(\sqrt{\rho_{\beta_i,\mu_i}} \; \overline{f}) \; ,\label{AW1} \\
\pi_i^\# (b_i (f)) &:=& \tilde{b}_i^* (\sqrt{\rho_{\beta_i,\mu_i}}f)(-1)^{N_i}\otimes (-1)^{N_i} +  \unit^{\R_i}\otimes (-1)^{N_i}\tilde{b}_i (\sqrt{1-\rho_{\beta_i,\mu_i }} \; \overline{f}) \; ,\nonumber
\end{eqnarray}
where $N_i=d\Gamma_i(1)$ is the particle number operator for
reservoir $\R_i.$ Furthermore, $\Omega^{\R_i}\otimes\Omega^{\R_i}\in\F_i^{AW}$ corresponds to the equilibrium
KMS state $\omega^{\R_i}$ of reservoir $\R_i.$

The free dynamics on the GNS Hilbert space $\F_i^{AW}$ of each reservoir $\R_i$ is generated by the
standard Liouvillean $\L^{\R_i}.$ The modular operator associated with $(\O^{\R_i},\omega^{\R_i})$ is given by
$$\Delta^{\R_i}=e^{-\beta_i \L^{\R_i}} \; ,$$
and the modular conjugation is given by
$$J^{\R_i}(\Psi\otimes\Phi)=(-1)^{N_i(N_i-1)/2}\overline{\Phi}\otimes (-1)^{N_i(N_i-1)/2}\overline{\Psi} ,$$
for $\Psi,\Phi\in \F_i^{AW};$ (see, for example, [BR]).


In order to apply the complex translation method developed in
[JP1,2,3], we map $\F_i^{AW}=\F^{\R_i}(L^2 ({\mathbf R}^+ ; \B))\otimes
\F^{\R_i}(L^2 ({\mathbf R}^+ ; \B))$ to $\F^{\R_i}(L^2
({\mathbf R} ; \B))$ using the isomorphism between $L^2 ({\mathbf
R}^+ ; \B)\oplus L^2 ({\mathbf R}^+ ; \B)$ and $L^2 ({\mathbf R}
; \B).$ To every $f\in L^2 ({\mathbf R}^+ ; \B),$ we associate functions
$f_{\beta,\mu}, f_{\beta,\mu}^\# \in L^2 ({\mathbf R};\B)$ by setting
\begin{equation}
f_{\beta,\mu} (u,\sigma) := \begin{cases}
\sqrt{m(u)}\sqrt{1-\rho_{\beta,\mu}(u)} f(u,\sigma) \; , & u\ge 0 \\
\sqrt{m(-u)}\sqrt{\rho_{\beta,\mu}(-u)} \; \overline{f}(-u,\sigma) \; , & u<0 \;
\end{cases} \; ,
\label{fglued}
\end{equation}
and
\begin{eqnarray}
f_{\beta,\mu}^{\#} (u,\sigma) &:=&
\begin{cases}
\sqrt{m(u)}i\sqrt{\rho_{\beta,\mu} (u)} f(u,\sigma),  u\ge 0 \\
\sqrt{m(-u)}i\sqrt{1-\rho_{\beta,\mu} (-u)} \; \overline{f}(-u,\sigma), u<0 \end{cases} \nonumber \\
&=& i\overline{f}_{\beta,\mu}(-u,\sigma) ,\label{fglued2}
\end{eqnarray}
where $m(u)du$ is the measure on ${\mathbf R}^+ ,$ see eq. \fer{AW1}. (For a discussion of this map, see Appendix A.)


Let $a_i$ and $a_i^*$ be the annihilation and creation operators on $\F^{\R_i}(L^2({\mathbf R},du; \B)).$ Then
\begin{align}
\pi_i(b^\#_i(f)) &\rightarrow a_i^\#(f_{\beta_i,\mu_i}), \\
\pi_i^\#(b^\#_i(f)) &\rightarrow i (-1)^{N_i} a^\#_i(f^\#_{\beta_i,\mu_i}) ;\\
\Omega^{\R_i}\otimes\Omega^{\R_i} &\rightarrow \tilde{\Omega}^{\R_i},
\end{align}
where $a_i^\#$ stands for $a_i$ or $a_i^*,$ and $\tilde{\Omega}^{\R_i}$ is the vacuum state in
$\F^{\R_i}(L^2({\mathbf R},\B)).$ Using eqs. \fer{twopoint} and \fer{AW1}, one readily verifies that
\begin{align*}
\langle \tilde{\Omega}^{\R_i},a^*_i(f_{\beta_i,\mu_i})a_i(f_{\beta_i,\mu_i})\tilde{\Omega}^{\R_i}\rangle &= \langle \tilde{\Omega}^{\R_i},a^*_i(f^\#_{\beta_i,\mu_i})a_i(f^\#_{\beta_i,\mu_i})\tilde{\Omega}^{\R_i}\rangle \\
&=\omega^{\R_i}(b_i^*(f)b_i(f))\\
&= (f,\rho_{\beta_i,\mu_i}(\cdot )f) \; .
\end{align*}

Moreover, the free Liouvillean on $\F^{\R_i}(L^2 ({\mathbf R};\B))$ for the reservoir $\R_i $ is mapped to
\begin{equation}
\L^{\R_i}=d\Gamma_i (u_i) \; ,
\end{equation}
where $u_i \in {\mathbf R}.$


\subsection{The coupled system}

The kinematical algebra of the total system, $\S\vee\R_1\vee\cdots \vee
\R_n,$ is given by
\begin{equation}
\O=\O^\S\otimes \O^{\R_1}\otimes \cdots \otimes\O^{\R_n} \; ,
\end{equation}
and the Heisenberg-picture dynamics of the uncoupled system is given by
\begin{equation}
\a_0^t = \a_\S^t \otimes \a_{\R_1}^t \otimes\cdots\otimes
\a_{\R_n}^t \; .
\end{equation}

The representation of $\O$ on
$\H:= \H^\S\otimes\H^\S\otimes\F^{\R_1}(L^2 ({\mathbf R};\B))\otimes\cdots \otimes\F^{\R_n}(L^2 ({\mathbf R};\B)),$
determined by the initial state
\begin{equation}
\omega = \omega^\S\otimes \omega^{\R_1}\otimes\cdots\otimes \omega^{\R_n}
\end{equation}
by the GNS construction, is given by
\begin{equation}
\pi=\pi_\S\otimes\pi_{\beta_1 }\otimes\cdots\otimes\pi_{\beta_n} , \\
\end{equation}
and an anti-representation commuting with $\pi$ by
\begin{equation}
\pi^\# =\pi_\S^\#\otimes\pi_{\beta_1 }^\# \otimes \cdots
\otimes\pi_{\beta_n }^\# \; .
\end{equation}
Moreover, let $\Omega:=\Omega^\S\otimes\tilde{\Omega}^\R_1\otimes\cdots\otimes\tilde{\Omega}^{\R_n}$ denote the vector in $\H$ corresponding to the state $\omega.$ Denote the double commutant of $\pi(\O)$ by $\M:=\pi(\O)'' ,$ which is the
smallest von Neumann algebra containing $\pi(\O)$.

The Liouvillean of the total uncoupled system is
given by
\begin{equation}
\L_0 = \L^\S +\sum_{i=1}^n \L^{\R_i}  \; .
\label{freeL}
\end{equation}
This defines a selfadjoint operator on $\H.$

For $a\in\O,$ we abreviate $\pi (a)$ by $a$ whenever
there is no danger of confusion. The modular operator of the total system is
$$\Delta = \Delta^\S\otimes\Delta^{\R_1}\otimes\cdots\otimes\Delta^{\R_n} ,$$
and the modular conjugation is
$$J=J^\S\otimes J^{\R_1} \otimes\cdots\otimes J^{\R_n} .$$
According to Tomita-Takesaki theory,
$$J\M J=\M'\; , \Delta^{it} \M \Delta^{-it}=\M \; ,$$
for $t\in {\mathbf R};$ (see for example [BR]). Furthermore, for $a\in\M,$
\begin{equation}
J\Delta^{1/2}a\Omega=a^*\Omega .
\label{MCP}
\end{equation}


The system $\S$ is coupled to the reservoirs $\R_1,\cdots ,\R_n,$ through an interaction $gV(t),$
where $V(t)\in\O$ is given by
\begin{equation}
V(t)=\sum_{i=1}^n \{ \sigma_- \otimes b_i^* (f_i (t)) + \sigma_+ \otimes b_i (f_i (t)) \} \; ,
\label{interactinghamiltonian}
\end{equation}
where $\sigma_\pm = \sigma_1 \pm i \sigma_2,$
and $f_i(t) \in L^2({\mathbf R}^+ ; \B), i=1,\cdots ,n,$
are form factors.\footnote{Note that this form of an interaction preserves the total number of fermions in $\S\vee\R_1\vee\cdots\vee\R_n,$ as required by gauge invariance of the first kind.}


The {\it standard} Liouvillean of the interacting system acting
on the GNS Hilbert space $\H$ is given by
\begin{equation}
\L_g (t) = \L_0 + gI(t)\; ,
\end{equation}
where the unperturbed Liouvillean is defined in \fer{freeL}, and the
interaction Liouvillean determined by the operator $V(t)$ is given by
\begin{align}
&I(t)=\{ V(t) - JV(t)J \} \nonumber\\
&=\sum_{i=1}^n \{ \sigma_-\otimes {\mathbf 1}^\S\otimes a_i^*(f_{i,
\beta_i, \mu_i}(t))+\sigma_+\otimes {\mathbf 1}^\S\otimes a_i(f_{i,
\beta_i, \mu_i}(t)) \nonumber \\
&- i {\mathbf 1}^\S\otimes \sigma_-\otimes (-1)^{N_i} a_i^*(f_{i,
\beta_i, \mu_i}^\#(t))-i {\mathbf 1}^\S\otimes \sigma_+\otimes
(-1)^{N_i} a_i(f_{i, \beta_i, \mu_i}^\# (t)) \} ,
\label{interaction}
\end{align}
where $a_i,a_i^*$ are the annihilation and creation operators on the fermionic Fock space $\F^{\R_i}(L^2({\mathbf R};\B)).$ Note that since the perturbation is bounded, the domain of $\L_g(t)$ is $\D(\L_g(t))=\D(\L_0).$

Let $\overline{U}_g$ be the propagator generated by the standard Liouvillean. It satisfies
\begin{equation}
\partial_t\overline{U}_g(t)=-i\L_g(t)\overline{U}_g(t)\; ; \overline{U}_g(0)=1 \; .
\end{equation}
The Heisenberg-picture evolution is given by
\begin{equation}
\label{StandEvol}
\a_g^t(a)=\overline{U}_g^*(t)a\overline{U}_g(t) \; ,
\end{equation}
for $a\in\O.$


Generally, the kernel of $\L_g(t), Ker \; \L_g,$ is expected to be empty when at least two of
the reservoirs have different temperatures; (see section 6 and [JP3,MMS1,2])\footnote{This is consistent with the fact that the coupled system is not expected to possess the property of return to equilibrium if the reservoirs have different temperatures (or chemical potentials). One can verify that, indeed, this is the case
when assumptions (A1)-(A3), below, are satisfied.}. This
motivates introducing the so called C-Liouvillean, $L_g$ , which
generates dynamics on a Banach space contained in $\H$ (isomorphic to
$\O$) and which, {\it by construction}, has a non-trivial kernel.

Consider the Banach space
$$\C(\O,\Omega):=\{ a\Omega : a\in \O \} ,$$
with norm $\| a\Omega \|_{\infty}=\|a\|.$ Since $\Omega$ is
separating for $\O,$ the norm $\|a\Omega\|_\infty$ is well-defined, and since $\Omega$ is cyclic for $\O,$
$\C(\O,\Omega)$ is dense in $\H.$

We set $\phi(a)=a\Omega,$ and define a propagator $U_g(t,t')$ by
\begin{equation}
\phi(\a_g^{t,t'} (a)) = U_g(t,t')a\Omega .
\label{floquetmap}
\end{equation}
Then
\begin{equation}
\partial_t U_g(t,t') = iL_g(t)U_g(t,t') \; {\mathrm with }\; U_g(t,t)=1 ,
\label{floquetpropagator}\\
\end{equation}
and
\begin{equation}
U_g(t',t)\Omega = \Omega .
\end{equation}

Differentiating \fer{floquetmap} with respect to $t,$ setting $t=t',$ and using \fer{floquetpropagator}, \fer{StandEvol} and \fer{MCP}, one obtains
\begin{align*}
[(\L_0+gV(t))a-a(\L_0+gV(t))]\Omega &= [(\L_0+gV(t))a-(V(t)a^*)^*]\Omega \\
&= (\L_0+gV(t)-gJ\Delta^{1/2}V(t)\Delta^{-1/2}J)a\Omega \\
&\equiv L_g(t)a\Omega \; .
\end{align*} Hence, the C-Liouvillean is given by
\begin{equation}
\label{CL} L_g(t):= \L_0 + gV(t) -
gJ\Delta^{1/2}V(t)\Delta^{-1/2}J \; .
\end{equation}
Note that, {\it by construction},
$$L_g(t)\Omega=0,$$
for all $t\in {\mathbf R}.$

Next, we discuss the assumptions on the interaction. For
$\delta>0,$ we define the strips in the complex plane
$$I(\delta):= \{ z\in {\mathbf C} : | \Im z | < \delta \}$$
and
\begin{equation}
I^- (\delta):= \{ z\in {\mathbf C} : -\delta < \Im z < 0 \} .
\label{Iminus}
\end{equation}
Moreover, for every function $f\in L^2({\mathbf R^+};\B),$
we define a function $\tilde{f}$ by setting
\begin{equation}
\tilde{f}:=
\begin{cases}
\sqrt{m(u)}f(u,\sigma) , u\ge 0  \\
\sqrt{m(|u|)} \; \overline{f}(|u|,\sigma) , u<0
\end{cases} ,\label{tildefi}
\end{equation}
where $m(u)du$ is the measure on ${\mathbf R}^+.$ Denote by $H^2(\delta , \B)$ the Hardy class of analytic
functions
$$h: I(\delta) \rightarrow \B , $$
with
$$\| h \|_{H^2 (\delta ,\B)} := \sup_{|\theta | < \delta} \int_{{\mathbf R}}\| h (u+i\theta )\|_\B^2 du < \infty . $$
We require the following basic assumptions on the interaction term.

\begin{itemize}

\item[(A1)] {\it Periodicity.}

The interaction term $V(t)$ is periodic with (a minimal) period $\tau < \infty$: $V(t)=V(t+\tau).$

\item[(A2)] {\it Regularity of the form factors.}

Assume that $\exists \delta >0 ,$ independent of $t$ and $i\in \{1, \cdots , n\} ,$ such that
$$e^{-\beta_i u /2 }f_{\beta_i,\mu_i} (u,t)\in H^2(\delta , \B ),$$
where $f_i$ is as in eq. \fer{interactinghamiltonian} and $f_{\beta,\mu}$ is defined in \fer{fglued}.


\item[(A3)] {\it Fermi Golden Rule.}

We assume that
$$\sum_{i=1}^n \| \tilde{f}_i (2\omega_0,t)\|^2_{\B} > 0,$$
where $\tilde{f}_i$ is defined in \fer{tildefi}.
which means that the coupling of $\S$ to the reservoirs is non-vanishing in second order perturbation theory.\footnote{For instance, when the reservoirs are formed of nonrelativistic fermions in ${\mathbf R}^3,$ an example of a form factor satisfying assumptions (A1)-(A3) is given by
$$f_i(u,t)=h_i(t)|u|^ne^{-|u|^2},$$
where $h_i(t)$ is a bounded, periodic function of $t\in{\mathbf R}$ and $n\ge 2.$}

\end{itemize}

Let $\hat{f}$ be the Fourier transform of $\tilde{f}$ given by
\begin{equation}
\hat{f}_m(u,\omega)=\frac{1}{\tau}\int_0^\tau dt e^{-im\omega t}\tilde{f}(u,t) ,
\end{equation}
where $\omega=\frac{2\pi}{\tau},$ and $\tau$ is the period of the interaction term (see assumption (A1)).
Then $\tilde{f}(u,t)=\sum_{m\in {\mathbf Z}}e^{i m\omega t}\hat{f}_m(u,\omega).$ It follows from (A2) and Parseval's theorem that
\begin{equation}
\sum_{m\in {\mathbf Z}}\| \hat{f}_m(u+m\omega,\omega)\|^2_\B < \infty ,
\end{equation}
for $u\in {\mathbf R}.$

\bigskip

Let $\tilde{U}_g$ be the propagator generated by the adjoint of
the C-Liouvillean, ie,
\begin{eqnarray}
\partial_t \tilde{U}_g (t,t')&=& -i\tilde{U}_g(t,t')L_g^* (t) ,\\
\tilde{U}_g (t,t) &=& 1 .
\end{eqnarray}
Assumption (A2) implies that the perturbation is bounded, and
hence the domain of $L_g^\# ,$ where $L_g^\# $ stands for $L_g$ or
$L_g^*,$ is
$$\D(L_g^\#) = \D(\L_0) ,$$
and $U_g,\tilde{U}_g$ are bounded and strongly continuous in $t$ and $t'.$


\section{The Floquet Liouvillean}

In this section, we extend Floquet theory for periodically driven
quantum systems at zero temperature to a theory for systems at
positive temperatures. The goal is to investigate whether the
state of the coupled system converges to a time-periodic state. We introduce the Floquet Liouvillean, which generates the dynamics on a
suitable Banach space, and we show in the following section that
the time-periodic state to which the state of the system
converges after very many periods is related to a {\it zero-energy
resonance} of the Floquet Liouvillean.

We consider the extended Hilbert space $\tilde{\H}:= L^2
([0,\tau])\otimes \H,$ where $\tau$ is the period of the
perturbation appearing in \fer{interactinghamiltonian} and
(A1), and we introduce the Floquet Liouvillean
\begin{equation}
K_g^* := -i\partial_t\otimes\unit + \unit\otimes L_g^*(t) ,
\end{equation}
with periodic boundary conditions in $t.$ Note that, under
assumption (A2), $K_g^*$ is a closed operator with domain $\D=\D(i\partial_t\otimes\unit)\cap\D(\unit\otimes\L_0).$

By Fourier transformation,
$\tilde{\H}$ is isomorphic to
$$\bigoplus_{n\in {\mathbf Z}} \langle e^{in\omega t}\rangle \otimes\H = \bigoplus_{n\in {\mathbf Z}} \h^{(n)}\otimes\H ,$$
where $\h^{(n)}:= \langle e^{in\omega t}\rangle$ and $\omega= \frac{2\pi}{\tau}.$

According to Floquet theory of quantum mechanical systems driven
by periodic perturbation [Ho,Ya1,Ya2], the semi-group generated by
$K_g^*$ is given by
\begin{equation}
\label{FSM} (e^{-i\sigma K_g^*} f)(t) =
\tilde{U}_g(t,t-\sigma)f(t-\sigma) \;,
\end{equation}
where $f \in \tilde{\H}$ and $\sigma\in {\mathbf R}$.
Relation (\ref{FSM}) can be seen by differentiating both sides
with respect to $\sigma$ and setting $\sigma=0$ (see [Ho]).
(Alternatively, one may use the Trotter product formula, [RS1].)

Note that if
\begin{equation}
K_g^* \phi=\lambda \phi \; ,
\end{equation}
for $\phi\equiv \phi (t) \in \tilde{\H}$ and $\lambda\in {\mathbf C}$, then
$\phi(t)$ satisfies
\begin{equation}
\tilde{U}_g(t,0)\phi(0)=e^{-i\lambda t}\phi(t) \; .
\end{equation}
Conversely, if
\begin{equation}
\tilde{U}_g(\tau,0)\phi (0)=e^{-i\lambda \tau}\phi_0 \; ,
\end{equation}
then
\begin{equation}
\phi(t)=e^{i\lambda t} \tilde{U}_g(t,0)\phi_0
\end{equation}
is an eigenfunction of $K_g^*$ with eigenvalue $\lambda$.


We now study the spectrum of $K_g^*$ using complex spectral deformation
techniques as developed in [HP,JP1,2,3].\footnote{One may alternatively use the methods developed in [MMS1,2].}

Let ${\mathbf u}_i$ be the unitary
transformation generating translations in energy for the $i^{th}$
reservoir, $i=1,\cdots,n.$ More precisely, for $f_i\in L^2({\mathbf R};\B),$
$${\mathbf u}_i(\theta) f_i (u) = f_i^{\theta}(u) = f_i (u+\theta)  .$$
Moreover, let
$$U_i(\theta):=\Gamma_i({\mathbf u}_i(\theta)) $$
denote the second quantization of ${\bf u}_i(\theta).$

Explicitly, $U_i(\theta)=e^{i\theta A_i},$ where $A_i := i
d\Gamma_i (\partial_{u_i})$ is the second quantization of the
generator of energy translations for the $i^{th}$reservoir,
$i=1,\cdots ,n.$ We set
\begin{equation}
U(\theta) := \unit^\S\otimes\unit^\S\otimes U_1(\theta)
\otimes\cdots\otimes U_n(\theta) \; .
\end{equation}

Let
\begin{align}
K^*_g(\theta) &:= U(\theta)K^*_g U(-\theta) \; \\
&= -i\partial_t + L_g^*(t,\theta) \; ,\label{deformedfloquet}
\end{align}
where $L^*_g(t,\theta )$ is given by
\begin{equation}
L^*_g (t, \theta):= U(\theta )L^*_g (t) U(-\theta) = \L_0 +
N\theta + g\tilde{V}^{tot}(t,\theta) \; , \label{CL}
\end{equation}
$\L_0=\L^\S+\sum_i \L^{\R_i}$, $\L^{\R_i}=d\Gamma (u_i),
i=1,\cdots ,n$, and
\begin{align*}
\tilde{V}^{tot}(t,\theta)&= \sum_i \{ \sigma_+\otimes {\mathbf
1}^\S \otimes a_i (f_{i,\beta_i, \mu_i}^{(\theta )}(t)) + \sigma_-\otimes\unit^\S \otimes a_i^* (f_{i,\beta_i,\mu_i}^{(\theta )}(t))\\&
- i\unit^\S \otimes \sigma_- \otimes (-1)^{N_i}(a_i(e^{\beta_i
(u_i-\mu_i)/2 }f_{i, \beta_i,\mu_i}^{\# (\theta )}(t))\\
& -i\unit^\S\otimes\sigma_+\otimes (-1)^{N_i} a_i^*(e^{-\beta_i
(u_i-\mu_i)/2}f_{i,\beta_i,\mu_i}^{\# (\theta )}(t)) \} \; .
\end{align*}

It follows from assuption (A2) that, for $\theta \in I(\delta),$
$\tilde{V}_g^{tot} (t,\theta)$ is a bounded operator. Hence $K^*_g(t,\theta)$ is
well-defined and closed on the domain $\D:=\D(i\partial_t)\cap\D(N)\cap\D(\L^{\R_1})\cap\cdots\cap \D(\L^{\R_n}).$ When the coupling $g=0$, the pure point spectrum of
$\L_0$ is $\sigma_{pp}(\L_0)=\{-2\omega_0,0,2\omega_0\}$,
with double degeneracy at 0, and the continuous spectrum of $\L_0$ is $\sigma_{cont}(\L_0)=\mathbf{R}$. It follows that
\begin{equation}
\label{K0pps} \sigma_{pp}(K_0)=\{ E_j^{(k)}(g=0)=E_j+k\omega :j=0,\cdots,3, k\in {\mathbf Z}\},
\end{equation}
where $E_{0,1}=0,E_2=-2\omega_0,$ and $E_3=2\omega_0,$ and $\sigma_{cont}(K_0)={\mathbf R}.$ Let $$K^\S:=-i\partial_t+\L^\S.$$
Clearly, $\sigma(K^\S)=\sigma_{pp}(K_0).$ We have the following two easy lemmas.

\bigskip
\noindent {\bf Lemma 4.1}

{\it For $\theta\in {\mathbf C}$, the following holds.
\begin{itemize}

\item[(i)] For any $\psi\in \D$, one has
\begin{equation}
\| K_0 (\theta ) \psi \|^2 = \| K_0 (\Re\theta) \psi \|^2 + |
\Im \theta |^2 \| N \psi \|^2 \; .
\end{equation}

\item[(ii)] If $\Im \theta \ne 0$, then $K_0 (\theta )$ is a normal operator satisfying
\begin{equation}
K_0(\theta)^*=K_0 (\overline{\theta}) \; ,
\end{equation}
and $\D(K_0(\theta))=\D$.

\item[(iii)] The spectrum of $K_0 (\theta )$ is
\begin{align}
\sigma_{cont} (K_0 (\theta )) &= \{ n\theta + s : n\in {\mathbf N\backslash\{0\} } \;
{\mathrm and} \;  s\in {\mathbf R} \} ,\\
\sigma_{pp}(K_0(\theta)) &= \{k\omega +E_j : j=0,\cdots,3,k\in {\mathbf Z}\} ,
\end{align}
where $E_{0,1}=0, E_2=-2\omega_0$ and $E_3=2\omega_0,$ (the
eigenvalues of $\L^\S$), and $\omega=\frac{2\pi}{\tau}.$
\end{itemize}
}
\bigskip

\noindent {\it Proof.} The first claim follows directly by looking at the
sector where $N=n\unit ,$ since $K_0 (\theta)$ restricted to this sector is
reduced to
\begin{equation}
K_0^{(n)}(\theta )=K^\S + s_1 + \cdots + s_n + n\theta\; ,\label{sector}
\end{equation}
where $s_1,\cdots,s_n$ are interpreted as one-particle multiplication operators.
For $\Im\theta\ne 0,$ it also follows from \fer{sector}
that
$$\D = \{ \psi = \{ \psi^{(n)} \} : \psi^{(n)} \in
\D(K_0^{(n)}(\theta)) \; {\mathrm and} \; \sum_n \|
K_0^{(n)}(\theta) \psi^{(n)}\|^2 < \infty \},$$ and hence
$K_0 (\theta )$ is a closed normal operator on $\D$. Claims (ii)
and (iii) follow from the corresponding statements on
$K_0^{(n)}(\theta )$. $\Box$

\bigskip

\noindent {\bf Lemma 4.2}

{\it Suppose (A1)-(A3) hold, and assume that $(g, \theta)\in {\mathbf
C}\times I^- (\delta ).$ Then the following holds.

\begin{itemize}

\item[(i)] $\D(K^*_g (\theta ))=\D$ and $(K^*_g (\theta))^*=K_{\overline{g}}(\overline{\theta})$.

\item[(ii)] The map $(g, \theta) \rightarrow K^*_g(\theta)$ from ${\mathbf C}\times I^- (\delta)$
to the set of closed operators on $\tilde{\H}$ is an analytic family (of type A) in each variable separately; (see [Ka1], chapter V, section 3.2).

\item[(iii)] For finite $g\in {\mathbf R}$ and $\Im z$ large enough,
\begin{equation}
s-\lim_{\Im \theta\uparrow 0}(K^*_g(\theta)-z)^{-1}= (K^*_g(\Re
\theta)-z)^{-1}\; . \label{removeCD2}
\end{equation}

\end{itemize}}

\bigskip

\noindent {\it Proof.} The first claim (i) follows from the fact that
$g\tilde{V}^{tot}(t, \theta)$ is bounded for $\theta\in I(\delta)$. It also follows from assumption (A2) that $(g, \theta)\rightarrow K^*_g(\theta)$ is analytic in $\theta\in I^-(\delta).$
Analyticity in $g$ is obvious from \fer{deformedfloquet}. We still need to prove claim $(iii).$ Without loss of generality, assume that $\Re \theta = 0$. It follows from assumption (A2) that the resolvent formula
\begin{equation}
\label{resolvent} (K^*_g (\theta)-z)^{-1} = (K^*_0 (\theta
)-z)^{-1}(1+ g\tilde{V}^{tot}(\cdot,\theta )(K^*_0 (\theta )-z)^{-1})^{-1} \; ,
\end{equation}
holds for small $g$, as long as $z$ belongs to the half-plane $\{ z\in
{\mathbf C}: 0< c< \Im z \}.$  Since $(K^*_0(\theta)-z)^{-1}$ is uniformly bounded as $\Im\theta\uparrow 0$ for $g\in {\mathbf R}$ finite and $\Im z$ large enough, and $\tilde{V}^{tot}(\theta)$ is bounded and analytic in $\theta$, claim (iii) follows from the Neumann series expansion of the resolvent of $K_g^*(\theta)$. $\Box$


\bigskip

Next, we apply degenerate perturbation theory, as developed in [HP], to compute the spectrum of $K_g^*(\theta).$ Using contour integration, one may define the projection onto the perturbed eigenstates of $K_g^*(\theta),$ for $\Im\theta\in I^-(\delta).$ Let
\begin{equation}
\tilde{P}_{g,\bk} (\theta):= \oint_{\gamma_k} \frac{dz}{2\pi i}(z-K^*_g (\theta))^{-1} \; ,
\end{equation}
where $\gamma_k$ is a contour that encloses the
eigenvalues $E_j^{\bk}(g=0),j=0,\cdots,3,k\in {\mathbf Z},$ at a distance $d>0,$ such that, for sufficiently small $|g|$ (to be specified below) the contour also encloses $E_j^{\bk}(g).$

Moreover, let $\tilde{T}_{g,\bk}:= \tilde{P}_{0,\bk} \tilde{P}_{g,\bk} (\theta)\tilde{P}_{0,\bk}.$\footnote{Note that, although $K_0^*(\theta)$ is unbounded, it is a normal operator, and hence $\tilde{P}_{0,\bk}$ is well-defined by the spectral theorem; (see for example [Ka1]).}
We show in Theorem 4.3 that the isomorphism
\begin{equation}
\label{quasiFL} \tilde{S}_{g,\bk}(\theta ):=
\tilde{T}_{g,\bk}^{-1/2} \tilde{P}_{0,\bk} \tilde{P}_{g,\bk}
(\theta): Ran(\tilde{P}_{g,\bk}(\theta))\rightarrow
\h^{(k)}\otimes\H^\S\otimes\H^\S
\end{equation}
has an inverse
\begin{equation}
\tilde{S}_{g,\bk}^{-1}(\theta):= \tilde{P}_{g,\bk}(\theta)
\tilde{P}_{0,\bk} \tilde{T}_{g,\bk}^{-1/2}(t):
\h^{(k)}\otimes\H^\S\otimes\H^\S\rightarrow
Ran(\tilde{P}_{g,\bk}(\theta)) .
\end{equation}
We set
\begin{equation}
\tilde{M}_{g,\bk}:= \tilde{P}_{0,\bk} \tilde{P}_{g,\bk}
(\theta)K^*_g(\theta)\tilde{P}_{g,\bk}(\theta)\tilde{P}_{0,\bk}\;,
\end{equation}
and define the quasi-Floquet Liouvillean by
\begin{equation}
\label{qFL} \tilde{\S}_{g,\bk} := \tilde{S}_{g,\bk}
(\theta)\tilde{P}_{g,\bk}(\theta)K^*_g(\theta)\tilde{P}_{g,\bk}(\theta)\tilde{S}_{g,\bk}^{-1}(\theta)=
\tilde{T}_{g,\bk}^{-1/2}\tilde{M}_{g,\bk}\tilde{T}_{g,\bk}^{-1/2}\; .
\end{equation}

Let $\kappa=min\{\delta,\frac{\pi}{\beta_1},\cdots,\frac{\pi}{\beta_n}\},$
where $\delta$ appears in assumption (A2), section 3, and $\beta_1,\cdots,\beta_n,$
are the inverse temperatures of the reservoirs $\R_1,\cdots,\R_n,$ respectively.
For $\theta \in I^-(\kappa)$ (see \fer{Iminus}), we choose a parameter $\nu$ such that
\begin{equation}
\label{nu}
-\kappa<\nu<0 \; \; {\mathrm and} \;  -\kappa<\Im\theta<-\frac{\kappa +|\nu|}{2}.
\end{equation}
We also choose a constant $g_1>0$ such that
\begin{equation}
g_1 C < (\kappa-|\nu|)/2,
\label{g1}
\end{equation}
where
\begin{equation}
C:=\sup_{\theta\in I(\delta),t\in {\mathbf R}}\| \tilde{V}^{tot}(t,\theta)\|<\infty.
\label{C}
\end{equation}

\bigskip
\noindent {\bf Theorem 4.3}

{\it Suppose that assumptions (A1)-(A3) hold. Then, for $g_1>0$ satisfying \fer{g1}, $\theta \in I^-(\kappa)$ and $\nu$ satisfying \fer{nu}, the following holds.

\begin{itemize}

\item[(i)] If $|g| < g_1$, the essential spectrum of the operator $K^*_g(\theta)$ is contained in the half-plane ${\mathbf C}\backslash \Xi(\nu),$ where $\Xi (\nu):= \{ z\in {\mathbf C} : \Im z \ge \nu \}.$ Moreover, the discrete spectrum of $K_g^*(\theta)$ is independent of $\theta\in I^-(\kappa)$. If $|g|< \frac{1}{2} g_1$, then the spectral projections $\tilde{P}_{g,\bk} (\theta), k\in {\mathbf Z},$
associated to the spectrum of $K^*_g (\theta)$ in the half-plane $\Xi (\nu),$ are analytic in $g$ and satisfy the estimate
\begin{equation}
\| \tilde{P}_{g,\bk}(\theta) - \tilde{P}_{0,\bk} \| < 1\; .
\end{equation}

\item[(ii)]If $|g|<\frac{g_1}{2}$, then the
quasi-Floquet Liouvillean $\tilde{\S}_{g,\bk}$ defined in
(\ref{qFL}) depends analytically on $g$, and has a Taylor
expansion
\begin{equation}
\label{FLTaylor2} \tilde{\S}_{g,\bk}=K^\S_\bk +
\sum_{j=1}^{\infty} g^{2j} \tilde{\S}_\bk^{(2j)} \;
\end{equation}
where
\begin{equation}
K^\S_\bk:= k\omega +\L^\S \;, k\in {\mathbf Z} .
\end{equation}

The first non-trivial coefficient in (\ref{FLTaylor2}) is
\begin{align*}
\tilde{\S}_\bk^{(2)}& = \frac{1}{2} \oint_\gamma \frac{dz}{2\pi
i}(\xi_\bk(z)(z-K_\bk^\S)^{-1}+(z-K_\bk^\S)^{-1}\xi_\bk(z))\; ,
\label{2orderqFL}
\end{align*}
where $\xi_\bk(z)=\tilde{P}_{0,\bk}
\tilde{V}^{tot}(\theta)(z-K_0(\theta))^{-1}\tilde{V}^{tot}(\theta)\tilde{P}_{0,\bk}.$

\end{itemize}}

\bigskip


\noindent {\it Proof.}
\noindent $(i)$ The resolvent formula
\begin{equation}
\label{resolvent} (K^*_g (\theta)-z)^{-1} = (K^*_0 (\theta
)-z)^{-1}(1+ g\tilde{V}^{tot}(\cdot,\theta )(K^*_0 (\theta )-z)^{-1})^{-1} \; ,
\end{equation}
holds for small $g$ and $z$ in the half-plane $\{ z\in
{\mathbf C}: 0< c< \Im z \}$. We extend the domain of validity of
(\ref{resolvent}) by refining the estimate on $g\tilde{V}^{tot}(t,
\theta)(K^*_0(\theta)-z)^{-1}.$

Note that
\begin{align*}
\| g\tilde{V}^{tot} (\cdot, \theta) (K^*_0 (\theta)-z)^{-1} \| &\le |g| C \| (K_0
(\theta)-z)^{-1} \|\\
&\le |g|C\frac{1}{dist(z,\eta(K_0^*(\theta)))},\label{ResEst}
\end{align*}
where $C$ is given by \fer{C} and $\eta(K_0^*(\theta))$ is the closure of the numerical range of $K_0^*.$ Fix $g_1$ such that it satisfies \fer{g1}, and choose $\epsilon$ such that $\epsilon> \frac{\kappa-|\nu|}{2} >0$. Let
$$G(\nu, \epsilon):= \{ z\in {\mathbf C}: \Im z > \nu
; dist(z,\eta (K^*_0 (\theta))>\epsilon\}.$$
Then
\begin{equation*}
\sup_{z\in G(\nu,\epsilon)}\| g \tilde{V}^{tot}(\cdot, \theta)(K^*_0
(\theta)-z)^{-1}\| \le \frac{|g|}{g_1}.
\end{equation*}
If $|g| < g_1$, the resolvent formula (\ref{resolvent})
holds on $G(\nu, \epsilon)$, and, for $m\ge 1$,
\begin{equation}
\sup_{z \in G( \nu , \epsilon )} \| (z-K^*_g (\theta))^{-1}-\sum_{j=0}^{m-1}(z-K^*_0 (\theta))^{-1}(g\tilde{V}^{tot}(\cdot ,
\theta)(z-K^*_0(\theta))^{-1})^j \|
\le \frac{(\frac{|g|}{g_1})^m}{1-\frac{|g|}{g_1}}.
\label{extendedresolvent}
\end{equation}
It follows that
\begin{equation}
\bigcup_{\epsilon>\frac{\kappa-|\nu|}{2}} G(\nu, \epsilon)\subset \rho (K^*_g(\theta)) \; ,
\end{equation}
where $\rho (K^*_g(\theta))$ is the resolvent set of $K^*_g (\theta)$.
Moreover, setting $m=1$ in \fer{extendedresolvent}, it follows that, for $|g|<g_1/2,$
$$\| \tilde{P}_{g,\bk}(\theta)-\tilde{P}_{0,\bk}\| < 1,$$
and hence $\tilde{P}_{g,\bk}$ is analytic in $g.$ We still
need to prove the independence of $\sigma_{pp}(K_g^*(\theta))$ of $\theta\in I^-(\kappa)$.

Fix $(g_0, \theta_0)\in {\mathbf C}\times I^- (\delta)$
such that $|g_0| < g_1$. The discrete eigenvalues
of $K^*_{g_0}(\theta)$ are analytic functions with at most
algebraic singularities in the neighbourhood of $\theta_0$, since
$K^*_{g_0}(\theta)$ is analytic in $\theta$. Moreover, since
$K^*_{g_0}(\theta_0)$ and $K^*_{g_0}(\theta)$ are unitarily
equivalent if $(\theta - \theta_0)\in {\mathbf R}$, it follows
that the pure point spectrum of $K^*_{g_0}(\theta)$ is independent
of $\theta$.

\noindent$(ii)$ Analyticity of $\tilde{T}_{g,\bk}$ directly follows from (i) and the definition of $\tilde{T}_{g,\bk}$.
Since $\| \tilde{T}_{g,\bk} - 1 \| < 1$ for $|g|< g_1 /2$,
$\tilde{T}_{g,\bk}^{-1/2}$ is also analytic in $g$. Inserting the Neumann series
for the resolvent of $K^*_g(\theta)$, gives
\begin{equation}
\tilde{T}_{g,\bk} = 1 + \sum_{j=1}^{\infty} g^j \tilde{T}_\bk^{(j)} \; ,
\end{equation}
with
\begin{equation}
\tilde{T}_\bk^{(j)} =\oint_{\gamma_k} \frac{dz}{2\pi i} (z-K^\S)^{-1} \tilde{P}_{0,\bk}
\tilde{V}^{tot}(\cdot, \theta) ((z-K^*_0(\theta))^{-1}\tilde{V}^{tot}(\cdot,
\theta))^{j-1}\tilde{P}_{0,\bk} (z-K^\S)^{-1}\; .
\end{equation}

Similarly,
\begin{equation}
\tilde{M}_{g,\bk}=K^\S+\sum_{j=1}^{\infty}g^j \tilde{M}_\bk^{(j)}\; ,
\end{equation}
with
\begin{equation}
\tilde{M}_\bk^{(j)}=\oint_\gamma \frac{dz}{2\pi i}z (z-K^\S)^{-1} \tilde{P}_{0,\bk}
\tilde{V}^{tot}(\cdot, \theta) ((z-K^*_0(\theta))^{-1}\tilde{V}^{tot}(\cdot,
\theta))^{j-1}\tilde{P}_{0,\bk} (z-K^\S)^{-1}\; .
\end{equation}

The odd terms in the above two expansions are zero due to the fact
that $\tilde{P}_{0,\bk}$ projects onto the $N=0$ sector. The first non-trivial
coefficient in the Taylor series of $\tilde{\S}_{g\bk}$ is
\begin{align}
\tilde{\S}^{(2)}_\bk &= \tilde{M}^{(2)}_\bk-\frac{1}{2} (\tilde{T}^{(2)}_\bk K^\S + K^\S \tilde{T}^{(2)}_\bk) \\
&= \frac{1}{2} \oint_{\gamma_k} \frac{dz}{2\pi
i}(\xi_k(z)(z-K^\S)^{-1}+(z-K^\S)^{-1}\xi_k(z))\; , \label{2orderqL}
\end{align}
with
$$\xi_k(z)=\tilde{P}_{0\bk} \tilde{V}_g^{tot}(\cdot,
\theta)(z-K^*_0(\theta))^{-1}\tilde{V}_g^{tot}(\cdot, \theta)\tilde{P}_{0\bk}.$$ $\Box$

\bigskip


We explicitly compute the discrete spectrum of
$K_g^*(\theta)$ to second order in the coupling constant. Let $e_{1,2}\in\H^\S$ denote the vectors of spin up and down respectively. Then the states in $\tilde{\H}$ corresponding to the eigenvalues $E^\bk_j(g=0),j=0,\cdots,3,k\in {\mathbf Z},$ are
\begin{align*}
\phi_k^0 &= e^{ik\omega}\otimes e_1\otimes e_1\otimes \tilde{\Omega}^{\R_1}\otimes\cdots\otimes\tilde{\Omega}^{\R_n} ,\\
\phi_k^1 &= e^{ik\omega}\otimes e_2\otimes e_2\otimes \tilde{\Omega}^{\R_1}\otimes\cdots\otimes\tilde{\Omega}^{\R_n} ,\\
\phi_k^2 &= e^{ik\omega}\otimes e_2\otimes e_1\otimes \tilde{\Omega}^{\R_1}\otimes\cdots\otimes\tilde{\Omega}^{\R_n} ,\\
\phi_k^3 &= e^{ik\omega}\otimes e_1\otimes e_2\otimes \tilde{\Omega}^{\R_1}\otimes\cdots\otimes\tilde{\Omega}^{\R_n} ,\\
\end{align*}
where $\Omega^{\R_i}$ is the vacuum state in $\F^{\R_i}(L^2({\mathbf R};\B)).$

We apply perturbation theory to calculate $E^\bk_j(g).$ We know that
\begin{eqnarray}
\tilde{\S}_\bk^{(2)} &=& \frac{1}{2}\oint_{\gamma_k} \frac{dz}{2\pi i} \{ \tilde{P}_{0,\bk} \tilde{V}^{tot}(\cdot,\theta )(z-K_0(\theta ))^{-1}\tilde{V}^{tot}(\cdot,\theta )\tilde{P}_{0,\bk} (z-K_\bk^\S)^{-1} \nonumber\\
&+&(z-K_\bk^\S) \tilde{P}_{0,\bk} \tilde{V}^{tot}(\cdot,\theta
)(z-K_0(\theta ))^{-1}\tilde{V}^{tot}(\cdot,\theta )\tilde{P}_{0,\bk} \}
\; .\label{gammak}
\end{eqnarray}
For $f_{\beta,\mu}$ as in \fer{fglued}, we let its Fourier transform be
\begin{equation}
\hat{f}_{\beta,\mu,m}(u,\omega):=\frac{1}{\tau}\int_0^\tau dt e^{-im\omega t} f_{\beta,\mu}(u,t) .
\end{equation}
Similarly, for $f_{\beta,\mu}^\#$ as in \fer{fglued2}, we let
\begin{equation}
\hat{f}^\#_{\beta,\mu,m}(u,\omega):=\frac{1}{\tau}\int_0^\tau dt e^{-im\omega t} f^\#_{\beta,\mu}(u,t) .
\end{equation}

Consider first the nondegenerate eigenvalue $E^\bk_3.$ Applying the Cauchy integration formula to \fer{gammak}, and using the facts that
\begin{equation*}
\lim_{\epsilon\searrow 0}\Re \frac{1}{x-i\epsilon} = \PV \frac{1}{x} ,\; {\mathrm and } \; \lim_{\epsilon\searrow 0}\Im \frac{1}{x-i\epsilon}= \pi\delta (x) ,
\end{equation*}
where $\PV$ denotes the Cauchy principal value, it follows that
\begin{eqnarray*}
\Re \ll \phi_k^3, \tilde{\S}^{k(2)}_3 \phi_k^3 \gg &=& \sum_{m\in {\mathbf Z}}\sum_{i=1}^n \PV \int_{{\mathbf R}} du \frac{\| \hat{f}_{\beta_i, \mu_i,m}(u,\omega)\|^2_\B}{2\omega_0-(k-m)\omega-u} \; , \\
\Im \ll \phi_k^3, \tilde{\S}^{k(2)}_3 \phi_k^3 \gg &=& -\pi\sum_{m\in {\mathbf Z}}\sum_{i=1}^n \| \hat{f}_{\beta_i, \mu_i,m}(2\omega_0-(k-m)\omega,\omega)\|^2_\B \; ,
\end{eqnarray*}
where $\ll\cdot , \cdot\gg$ is the scalar product on $\tilde{\H}.$
Therefore,
\begin{align*}
E^\bk_3(g)&=k\omega+2\omega_0 +g^2 \sum_{m\in {\mathbf Z}}\sum_{i=1}^n \PV \int_{{\mathbf R}} du \frac{\| \hat{f}_{\beta_i, \mu_i,m}(u,\omega)\|^2_\B}{2\omega_0-(k-m)\omega-u} -\\
&-i\pi g^2\sum_{m\in {\mathbf Z}}\sum_{i=1}^n \| \hat{f}_{\beta_i, \mu_i,m}(2\omega_0-(k-m)\omega,\omega)\|^2_\B + \O(g^4) .
\end{align*}
Similarly,
\begin{align*}
E^\bk_2(g)&=k\omega-2\omega_0 -g^2 \sum_{m\in {\mathbf Z}}\sum_{i=1}^n \PV \int_{{\mathbf R}} du \frac{\| \hat{f}_{\beta_i, \mu_i,m}(u,\omega)\|^2_\B}{2\omega_0-(k-m)\omega-u} -\\ &-i\pi g^2\sum_{m\in {\mathbf Z}}\sum_{i=1}^n \| \hat{f}_{\beta_i, \mu_i,m}(2\omega_0-(k-m)\omega,\omega)\|^2_\B + \O(g^4) .
\end{align*}

Next we use degenerate perturbation theory to calculate $E^\bk_j(g),j=0,1.$ Applying the Cauchy integration formula to \fer{gammak} and using the definitions of $f_{\beta,\mu}$ and $f^\#_{\beta,\mu}$, it follows that
\begin{align*}
Re \ll \phi_k^{0,1}, \tilde{\S}^{k(2)}_3 \phi_k^{0,1} \gg &= -Re \ll \phi_k^{1,0}, \tilde{\S}^{k(2)}_3 \phi_k^{0,1} \gg \\&= \pm \sum_{m\in {\mathbf Z}}\sum_{i=1}^n \PV \int_{{\mathbf R}} du \frac{\| \hat{f}_{\beta_i, \mu_i,m}(u,\omega)\|^2_\B}{2\omega_0-(k-m)\omega-u} \; , \\
\Im \ll \phi_k^{0,1}, \tilde{\S}^{k(2)}_3 \phi_k^{0,1} \gg &=-\Im \ll \phi_k^{1,0}, \tilde{\S}^{k(2)}_3 \phi_k^{0,1} \gg \\&= -\pi\sum_{m\in {\mathbf Z}}\sum_{i=1}^n \| \hat{f}_{\beta_i, \mu_i,m}(2\omega_0-(k-m)\omega,\omega)\|^2_\B .
\end{align*}
Therefore,
\begin{equation}
E^{(k)}_{0,1}(g)=k\omega+ g^2 a_{0,1} + O(g^4)\; ,
\end{equation}
where $a_{0,1}$ are the eigenvalues of the $2\times 2$ matrix
\begin{equation}
-i \pi \sum_{m\in{\mathbf Z}}\sum_{i=1}^n \| \hat{f}_{\beta_i, \mu_i,m}(2\omega_0-(k-m)\omega,\omega)\|^2_\B \left(
\begin{matrix}
1  & -1  \\
-1 & 1
\end{matrix}
\right) \; .
\end{equation}
By construction, $K_g e^{i k\omega t}\Omega=k\omega e^{ik\omega
t}\Omega$ and $U(\theta)e^{ik\omega t}\Omega= e^{ik\omega t
}\Omega$, so $\{ k\omega : k\in\mathbf{Z}\}$ are also isolated
eigenvalues of $K_g^*(\theta), \theta\in I^-(\delta).$ This can be seen by
defining the spectral projections corresponding to the real isolated eigenvalues
of $K_g(\theta),$ using the resolvent, and taking the adjoint to
define the corresponding spectral projections for the real
isolated eigenvalues of $K_g^*(\theta).$\footnote{Suppose $\a$ is an isolated and real eigenvalue of a closed operator $A.$ Then the spectral projection corresponding to $\a$ is
$$P=\frac{1}{2\pi i}\oint_{\gamma_\a}(z-A)^{-1}dz,$$ where $\gamma_\a$ is a contour enclosing $\a$ only. Since $\a$ is real and isolated, it is also an eigenvalue of $A^*$ with corresponding projection $P^*.$ (Using Cauchy's integration formula, one may readily verify that $A^*P^*=P^*A^*=\a P^*$ and that $(P^*)^2=P^*.$)}

The vector $\psi=\left( \begin{matrix} 1
\\ 1
\end{matrix}\right)$ is the eigenvector corresponding the eigenvalue 0 of $\tilde{\S}_{g,\bk}^{2}$. Hence,
\begin{align}
E^{(k)}_0(g)&=k\omega \; ,\\
E^{(k)}_1(g)&=k\omega -2\pi i g^2 \sum_{m\in {\mathbf Z}}\sum_{i=1}^n \| \hat{f}_{\beta_i,\mu_i,m} (2\omega_0-(k-m)\omega)
\|^2_\B +O(g^4) \; .
\end{align}
Note that due to assumption (A3), $\Im E^{(k)}_j <0,$ for $j=1,2,3,$ while $\Im E^\bk_0 = 0.$\footnote{Alternatively, one can use the Feshbach map (see [BFS]) to compute the perturbation of the discrete spectrum of $K_g^*(\theta),$ which gives the same result.}
We have proven the following corollary to Theorem 4.3.

\begin{center}
\begin{picture}(0,0)%
\includegraphics{spectrum1.pstex}%
\end{picture}%
\setlength{\unitlength}{3729sp}%
\begingroup\makeatletter\ifx\SetFigFont\undefined%
\gdef\SetFigFont#1#2#3#4#5{%
  \reset@font\fontsize{#1}{#2pt}%
  \fontfamily{#3}\fontseries{#4}\fontshape{#5}%
  \selectfont}%
\fi\endgroup%
\begin{picture}(6144,3006)(3724,-4324)
\put(4366,-2446){\makebox(0,0)[lb]{\smash{{\SetFigFont{11}{13.2}{\familydefault}{\mddefault}{\updefault}{\color[rgb]{0,0,0}$k$}%
}}}}
\put(9001,-1501){\makebox(0,0)[lb]{\smash{{\SetFigFont{11}{13.2}{\familydefault}{\mddefault}{\updefault}{\color[rgb]{0,0,0}$2\omega$}%
}}}}
\put(5446,-1501){\makebox(0,0)[lb]{\smash{{\SetFigFont{11}{13.2}{\familydefault}{\mddefault}{\updefault}{\color[rgb]{0,0,0}-$\omega$}%
}}}}
\put(4321,-1501){\makebox(0,0)[lb]{\smash{{\SetFigFont{11}{13.2}{\familydefault}{\mddefault}{\updefault}{\color[rgb]{0,0,0}-2$\omega$}%
}}}}
\put(5581,-4246){\makebox(0,0)[lb]{\smash{{\SetFigFont{11}{13.2}{\familydefault}{\mddefault}{\updefault}{\color[rgb]{0,0,0}Figure 1. Spectrum of $K_g^*(\theta)$}%
}}}}
\put(6166,-1501){\makebox(0,0)[lb]{\smash{{\SetFigFont{11}{13.2}{\familydefault}{\mddefault}{\updefault}{\color[rgb]{0,0,0}-2$\omega_0$}%
}}}}
\put(7066,-1501){\makebox(0,0)[lb]{\smash{{\SetFigFont{11}{13.2}{\familydefault}{\mddefault}{\updefault}{\color[rgb]{0,0,0}2$\omega_0$}%
}}}}
\put(7876,-1501){\makebox(0,0)[lb]{\smash{{\SetFigFont{11}{13.2}{\familydefault}{\mddefault}{\updefault}{\color[rgb]{0,0,0}$\omega$}%
}}}}
\put(5041,-2446){\makebox(0,0)[lb]{\smash{{\SetFigFont{11}{13.2}{\familydefault}{\mddefault}{\updefault}{\color[rgb]{0,0,0}-$Im\theta$}%
}}}}
\end{picture}%

\end{center}

\bigskip

\bigskip
\noindent {\bf Corollary 4.4}

{\it Suppose assumptions (A1)-(A3) hold. Then, for $\theta\in I^-(\kappa)$ and $|g|<g_1/2,$ where $g_1$
satisfies \fer{g1}, $K_g^*(\theta)$ has infinitely many simple eigenvalues, $\{ k\omega\}_{k\in {\mathbf Z}},$ on the real axis, where $\omega=\frac{2\pi}{\tau}.$}
\bigskip

We will use the results of Theorem 4.3 and Corollary 4.4 to prove that the true state of the coupled system converges to a time-periodic state.


\section{Convergence to time-periodic states}

The following theorem claims that, under suitable
assumptions, the true state of the system converges to a
time-periodic state with the period $\tau$ of the perturbation.

Choose $\kappa=\min(\delta,\frac{\pi}{\beta_1},\cdots,\frac{\pi}{\beta_n})$ (as in section 4), define $\h^{test}:= D(e^{\kappa\sqrt{p^2+1}})$, and let $\O^{test,\R}$ be the $*$-algebra generated by $b^\#(f),f\in \h^{test},$ and by $\unit^{\R}$. Note that
$\O^{test,\R}$ is norm-dense in $\O^\R.$ We define a $*$-algebra
\begin{equation}
{\mathcal C}:= \O^\S\otimes\O^{\R_1 ,test}\otimes \cdots \otimes
\O^{\R_n, test} \; ,
\end{equation}
which is dense in $\O$.

We make the following additional assumption.
\begin{itemize}
\item[(A4)] The interaction Hamiltonian $V(t)$ belongs to ${\mathcal C},$ for $t\in {\mathbf R}$.

\end{itemize}

Let
\begin{equation}
D:=\unit^\S\otimes\unit^\S\otimes
e^{-\kappa\sqrt{A_{\R_1}^2+1}}\otimes\cdots\otimes
e^{-\kappa\sqrt{A_{\R_n}^2+1}} \; ,
\label{D}
\end{equation}
where $A_{\R_i}=d\Gamma(i\partial_{u_i}), i=1,\cdots,n,$ is the
second quantization of the generator of energy translations for
the $i^{th}$ reservoir. The operator $D$ is positive such that $D\Omega=\Omega$, and $Ran D$ is dense in $\H$. This opertor will be used as a regulator in order to apply the complex deformation technique. We have the following theorem. We will show in the following theorem that the time-periodic state to which the real state of the coupled system converges is related to a zero-energy resonance state given by $\tilde{\Omega}_{g,(0)}:= (\unit\otimes D)
\tilde{P}_{g,(0)}(\unit\otimes\Omega) \; .$

\bigskip

\noindent {\bf Theorem 5.1 (Convergence to time-periodic states).}

{\it Assume assumptions (A1)-(A4) hold. Assume
further that $a\in {\mathcal C}$. Then there is a constant
$g_1>0$ satisfying \fer{g1}, such that, for $|g|<g_1/2$, the following holds
\begin{equation}
\lim_{n\rightarrow\infty} \langle \Omega, \a_g^{n\tau+t}(a)
\Omega \rangle = \langle \tilde{\Omega}_{g,(0)} , D^{-1} \a_g^t (a)
\Omega \rangle \; ,
\end{equation}
where $\tilde{\Omega}_{g,(0)}$ corresponds to the zero-energy resonance
of the adjoint of the Floquet Liouvillean, $K_g^*$, and $D$ is given by \fer{D}.}

\bigskip

\noindent {\it Proof.} First note that by using a Dyson series expansion, it follows from
assumption (A4) and the fact that $a\in {\mathcal C}$ that $\a_g^t
(a)\in \C,$ and hence $\a_g^t(a)\Omega\in \D(D^{-1}).$

The remainder of the proof relies on the result of Theorem 4.3, Corollary 4.4,
and equation \fer{FSM} (section 4). It follows from \fer{FSM} and the time
periodicity of $f\in\tilde{\H}=L^2([0,\tau])\otimes\H$ that
\begin{equation}
\label{evolK} (e^{-iK_g^* n \tau}
\unit\otimes\Omega)(0)=\tilde{U}_g(n\tau,0)
(\unit\otimes\Omega)(0)=\tilde{U}_g(n\tau,0)\Omega \; .
\end{equation}
Let $\unit\otimes\Omega=: \overline{\Omega}\in \tilde{\H},$ and
$\overline{D}:=\unit\otimes D.$

Without loss of generality, we assume that $\omega\equiv\frac{2\pi}{\tau}\ne 2\omega_0;$
 (if $\frac{2\pi}{\tau}=2\omega_0$, the state of the system
typically oscillates between two resonance states until it finally
converges to a time-periodic state; see remark 1). Using the
dynamics on $C(\O,\Omega)$ and \fer{evolK}, it follows that
\begin{align}
&\lim_{n\rightarrow\infty}\langle \Omega, \a_g^{n\tau+t}(a) \Omega \rangle
= \lim_{n\rightarrow\infty}\langle \tilde{U}_g(n\tau, 0)\Omega, \a_g^t(a)\Omega\rangle \\
&= \lim_{n\rightarrow\infty} \langle (e^{-iK_g^* n\tau} \unitOmega)(0),\a_g^{t}(a)\Omega\rangle
\end{align}
Using the regulator $\overline{D}$ and complex spectral translation,
\begin{align}
&\lim_{n\rightarrow\infty}\langle \Omega, \a_g^{n\tau+t}(a) \Omega \rangle
= \lim_{n\rightarrow\infty} \langle (\unitD
U(-\theta)e^{-iK^*_g(\theta)n \tau}U(\theta) \unitD \;
\unitOmega)(0), D^{-1} \a^t_g(a) \Omega \rangle\\
&= \lim_{n\rightarrow\infty} \langle (\unitD
U(-\theta)\int_{-\infty}^\infty du (u+i\eta-K_g^*(\theta
))^{-1}e^{-i(u+i\eta)n\tau} \nonumber \\
&\unitOmega)(0),D^{-1} \a_g^t(a)\Omega \rangle .
\end{align}

We split the above integration into two terms,
\begin{align}
&\lim_{n\rightarrow\infty} \langle (\unitD
U(-\theta)\int_{-\infty}^\infty du (u+i\eta-K_g^*(\theta
))^{-1}e^{-i(u+i\eta)n\tau} \nonumber \\
&\unitOmega)(0),D^{-1} \a_g^t(a)\Omega \rangle\\
&= \lim_{n\rightarrow\infty} \langle  (\unitD U(-\theta)
\oint_\gamma dz (z -K_g^*(\theta))^{-1}e^{-iz n\tau } \unitOmega)(0) ,D^{-1} \a_g^t(a) \Omega \rangle + \nonumber \\
&+\lim_{n\rightarrow\infty} \langle (\unitD U (-\theta )
\int_{-\infty}^\infty du (u-i(\mu -\epsilon )-K_g^*(\theta
))^{-1}e^{-i(u-i(\mu-\epsilon ))n\tau}\unitOmega)(0), \nonumber \\
& D^{-1} \alpha_g^t(a) \Omega \rangle , \label{2term}
\end{align}
where $\eta>0, 0<\epsilon<\mu,$ and $\gamma$ is the contour
enclosing the point spectrum of $K_g^*(\theta)$ only.

Using the results of Theorem 4.3 and Corollary 4.4, the first term converges to a time-periodic expression,
\begin{align}
&\lim_{n\rightarrow\infty} \langle (\unitD U(-\theta) \oint_\gamma
dz (z-K_g^*(\theta))^{-1}e^{-iz n\tau}\unitOmega )(0),D^{-1}
\a_g^t(a)\Omega \rangle \\
&= \sum_{k\in{\mathbf Z}}\lim_{n\rightarrow\infty} \langle
(\unitD U(-\theta)\tilde{S}_{g,(k)}^{-1}(\theta)e^{-i\tilde{\S}_{g,(k)}(\theta)n\tau}\tilde{S}_{g,(k)}(\theta )\unitOmega )(0), D^{-1}\a_g^t(a)\Omega \rangle \\
&=\sum_{k\in {\mathbf Z}} \langle (\unitD
\tilde{P}_{g,(k)}(\theta) \unitOmega)(0),D^{-1} \a_g^t(a) \Omega\rangle
\end{align}

Let
\begin{equation}
\tilde{\Omega}_{g,(k)}:= \unitD
\tilde{P}_{g,(k)}(\theta)(e^{ik\omega t}\otimes\Omega) \; ,
\end{equation}
and denote by $\ll \cdot,\cdot \gg$ the scalar product on $\tilde{\H}.$
Then
\begin{equation}
\unitD\tilde{P}_{g,(k)}(\theta) \unitD = \ll (e^{ik\omega t}
\otimes\Omega), \cdot \gg \tilde{\Omega}_{g,(k)} \; .
\end{equation}

Therefore,
\begin{align}
\unitD \tilde{P}_{g,(k)}(\theta) \unitOmega
&= \ll e^{ik\omega t} \otimes \Omega, \unit\otimes\Omega \gg \tilde{\Omega}_{g,(k)} \\
&= \tilde{\Omega}_{g,(0)}\delta_{k,0} \; ,
\end{align}
where $\delta_{k,0}$ is the Kronecker delta, and hence
\begin{equation}
\sum_{k\in {\mathbf Z}} \langle (\unitD \tilde{P}_{g,(k)}
\unitOmega)(0),D^{-1}\a_g^t (a)\Omega \rangle =\langle
(\tilde{\Omega}_{g,(0)})(0) , D^{-1} \a_g^t (a) \Omega \rangle \; ,
\end{equation}
where $\tilde{\Omega}_{g,(0)}$ is the zero-energy resonance of the
Floquet Liouvillean. The second term in \fer{2term} converges
exponentially fast to zero since
\begin{equation}
\langle (\unitD U(-\theta) \int_{-\infty}^\infty du
(u-i(\mu-\epsilon)-K_g^*(\theta))^{-1}e^{-i(u-i(\mu-\epsilon))n\tau}
\unitOmega)(0),D^{-1} \a_g^t(a)\Omega\rangle =
O(e^{-(\mu-\epsilon')n\tau}) \; ,
\end{equation}
where $0<\epsilon'<\epsilon<\mu$; (see [Rud], chapter 19). $\Box$

\bigskip

\noindent {\it Remarks.}
\begin{itemize}

\item[(1)]When $\omega=\frac{2\pi}{\tau}=2\omega_0$, the system exhibits the
phenomenon of resonance: The state of the system oscillates between
two resonances until it finally converges to the time periodic
state corresponding to $\tilde{\Omega}_{g(0)}$. This can be
verified by a second order time-dependent perturbation theory
calculation; (see also [Ya2]).

\item[(2)] By a standard argument, the result of Theorem 5.1 can be extended to any initial state normal to $\omega$ (see for example [JP3,MMS1,2]).

\item[(3)]Note that
\begin{align*}
\langle (\tilde{\Omega}_{g,(0)})(0), D^{-1}\a^t_g(a)\Omega \rangle
= \langle (\tilde{\Omega}_{g,(k)})(0), D^{-1}\a^t_g(a)e^{ik\omega
t}\Omega\rangle \; ,
\end{align*}
where $\tilde{\Omega}_{g,(k)}$ is the state corresponding to the
$k\omega$-energy resonance of the adjoint of the Floquet
Liouvillean. In other words, all $k\omega$-energy resonances belong to the same class of time-periodic states.

\end{itemize}

In the next section, we discuss the positivity of entropy production per cycle and Carnot's formulation of the second law of thermodynamics.

\section{Positivity of entropy production}

We consider a small system coupled to two fermionic reservoirs at the same chemical potential $\mu$, yet at two different temperatures $\beta_1$ and $\beta_2$, with $\beta_1<\beta_2.$ Together with assumptions (A1)-(A4), we assume that the perturbation is differentiable in $t,$ for $t>0.$ The first reservoir acts as a heat source, and the second reservoir as
a heat sink. We want to show that, after the true state of the system has converged to a time-periodic state, the entropy
production per cycle is {\it strictly} positive. We first prove that the time-periodic state, which the true state of the system converges to, is not normal to the initial state. \footnote{In this section, we will use standard results about von Neumann algebras and Tomita-Takesaki modular theory. We refer the reader to [BR], chapters 2.4 and 2.5 for an exposition of these results.}

We introduce the {\it standard} Floquet Liouvillean,
\begin{equation}
\label{standardfloquet}
\tilde{K}_g:=-i\partial_t+\L_g(t) \; ,
\end{equation}
acting on the extended Hilbert space $L^2([0,\tau])\otimes \H,$ with periodic boundary conditions in $t$, where $\L_g(t)=\L_0+gV(t)-gJV(t)J,$ is the standard Liouvillean. We study the spectrum of the standard Floquet Liouvillean using complex spectral translations. Since the proof of the following proposition is very similar to the analysis in section 4, we only sketch the main steps of the proof.

\bigskip
\noindent {\bf Proposition 6.1}

{\it Suppose assumptions (A1)-(A3) (section 3) hold. Then there exists a positive constant $g_2$, such that, for $|g|<g_2,$ the spectrum of the standard Floquet Liouvillean $\tilde{K}_g$ defined in \fer{standardfloquet}, $\sigma(\tilde{K}_g),$  is absolutely continuous and $$\sigma (\tilde{K}_g)=\sigma_{ac} (\tilde{K}_g)={\mathbf R}.$$}

\bigskip

\noindent{\it Sketch of proof.}

Let $U(\theta)$ as in section 4. We define the complex deformed standard Floquet Liouvillean by

\begin{align}
\tilde{K_g}(t,\theta)&:=U(\theta)\tilde{K}_gU(-\theta)\\
&= -i\partial_t +\L_g(t,\theta) \; ,
\end{align}
where
$$\L_g(t,\theta)=\L_0+\theta N + gV^{tot}(t,\theta) \; ,$$
and
\begin{align*}
V^{tot}(t,\theta)&=U(\theta)(V(t)-JV(t)J)U(-\theta)\\
&=\sum_{i=1}^n \{ \sigma_-\otimes {\mathbf 1}^\S\otimes a^*(f_{i,
\beta_i,\mu}^{(\theta )}(t))+\sigma_+\otimes {\mathbf 1}^\S\otimes a(f_{i,
\beta_i,\mu}^{(\theta )}(t))  \\
&- i {\mathbf 1}^\S\otimes \sigma_-\otimes (-1)^{N_i} a^*(f_{i, \beta_i,\mu}^{\# (\theta )}(t))-i {\mathbf 1}^\S\otimes \sigma_+\otimes (-1)^{N_i} a(f_{i, \beta_i,\mu}^{\# (\theta)}(t))\} \; .
\end{align*}
It follows from assumption (A2) that $V^{tot}(t,\theta)$ is bounded for $\theta\in I(\delta).$ Let
\begin{equation}
\tilde{C}:=\sup_{t\in {\mathbf R},\theta\in I(\delta)}\|V^{tot}(t,\theta)\| .
\label{tildeC}
\end{equation}
For $\theta\in I^-(\kappa),$ choose
$\nu$ such that $0>\nu>-\kappa$ and $-\kappa<\Im\theta <-(\kappa+|\nu|)/2.$ Choose $g_2>0$ such that
\begin{equation}
g_2\tilde{C}<(\kappa+\nu)/4.
\label{g2}
\end{equation}
Then, using an argument which is similar to the proof of Theorem 4.3, one can show that, for $|g|<g_2,$ the essential spectrum of $\tilde{K}_g(\theta)$ is contained in the half-plane $\{z\in {\mathbf C}: \Im z<\nu\},$ and that it discrete spectrum
$$\sigma_{pp}(\tilde{K}_g)=\{ \tilde{E}^\bk_j(g) : k\in {\mathbf Z}, j=0,\cdots,3\} ,$$
where (to second order in perturbation theory)
\begin{align*}
\tilde{E}^\bk_{2,3}(g)&=k\omega\mp 2\omega_0 \mp g^2 \sum_{m\in {\mathbf Z}}\sum_{i=1}^2 \PV \int_{{\mathbf R}} du \frac{\| \hat{f}_{\beta_i, \mu_i,m}(u,\omega)\|^2_\B}{2\omega_0-(k-m)\omega-u} -\\
&-i\pi g^2\sum_{m\in {\mathbf Z}}\sum_{i=1}^2 \| \hat{f}_{\beta_i, \mu_i,m}(2\omega_0-(k-m)\omega,\omega)\|^2_\B + \O(g^4) ,
\end{align*}
while
\begin{equation}
\tilde{E}^\bk_{0,1}(g)=k\omega + g^2 a_{0,1} + O(g^4)\; ,
\end{equation}
where $a_{0,1}$ are the eigenvalues of the $2\times 2$ matrix
\begin{equation}
-i \pi \sum_{m\in{\mathbf Z}}\sum_{i=1}^2 \| \hat{f}_{\beta_i, \mu_i,m}(2\omega_0-(k-m)\omega,\omega)\|^2_\B \left(
\begin{matrix}
1  & -e^{-\beta_i(2\omega_0-(k-m)\omega-\mu)/2}  \\
-e^{\beta_i(2\omega_0-(k-m)\omega-\mu)/2} & 1
\end{matrix}
\right) \; .
\end{equation}
Note that, to second order in the coupling $g,$ the discrete spectrum is below the real axis.
The claim of the theorem follows by noting that
$$s-\lim_{\Im\theta\uparrow 0}(\tilde{K}_g(\theta)-z)^{-1}=(\tilde{K}_g(\Re\theta)-z)^{-1}$$
for small real $g$ and large enough $\Im z.$ $\Box$

\begin{center}
\begin{picture}(0,0)%
\includegraphics{spectrum2.pstex}%
\end{picture}%
\setlength{\unitlength}{3729sp}%
\begingroup\makeatletter\ifx\SetFigFont\undefined%
\gdef\SetFigFont#1#2#3#4#5{%
  \reset@font\fontsize{#1}{#2pt}%
  \fontfamily{#3}\fontseries{#4}\fontshape{#5}%
  \selectfont}%
\fi\endgroup%
\begin{picture}(6099,3051)(3769,-4369)
\put(4366,-2446){\makebox(0,0)[lb]{\smash{{\SetFigFont{11}{13.2}{\familydefault}{\mddefault}{\updefault}{\color[rgb]{0,0,0}$k$}%
}}}}
\put(9001,-1501){\makebox(0,0)[lb]{\smash{{\SetFigFont{11}{13.2}{\familydefault}{\mddefault}{\updefault}{\color[rgb]{0,0,0}$2\omega$}%
}}}}
\put(5446,-1501){\makebox(0,0)[lb]{\smash{{\SetFigFont{11}{13.2}{\familydefault}{\mddefault}{\updefault}{\color[rgb]{0,0,0}-$\omega$}%
}}}}
\put(4321,-1501){\makebox(0,0)[lb]{\smash{{\SetFigFont{11}{13.2}{\familydefault}{\mddefault}{\updefault}{\color[rgb]{0,0,0}-2$\omega$}%
}}}}
\put(5356,-4291){\makebox(0,0)[lb]{\smash{{\SetFigFont{11}{13.2}{\familydefault}{\mddefault}{\updefault}{\color[rgb]{0,0,0}Figure 2. Spectrum of  $\tilde{K}_g(\theta)$}%
}}}}
\put(6121,-1501){\makebox(0,0)[lb]{\smash{{\SetFigFont{11}{13.2}{\familydefault}{\mddefault}{\updefault}{\color[rgb]{0,0,0}-2$\omega_0$}%
}}}}
\put(7111,-1501){\makebox(0,0)[lb]{\smash{{\SetFigFont{11}{13.2}{\familydefault}{\mddefault}{\updefault}{\color[rgb]{0,0,0}2$\omega_0$}%
}}}}
\put(7876,-1501){\makebox(0,0)[lb]{\smash{{\SetFigFont{11}{13.2}{\familydefault}{\mddefault}{\updefault}{\color[rgb]{0,0,0}$\omega$}%
}}}}
\put(5041,-2446){\makebox(0,0)[lb]{\smash{{\SetFigFont{11}{13.2}{\familydefault}{\mddefault}{\updefault}{\color[rgb]{0,0,0}-$Im\theta$}%
}}}}
\end{picture}%

\end{center}


This result is sufficient to show that the time periodic state $\omega_{g,s}^+$ defined in \fer{timeper}, section 2, is not normal to the initial state $\omega.$ Let $g_3:=\min\{g_1/2,g_2\},$ where $g_1$ satisfies \fer{g1} (section 4) and $g_2$ satisfies \fer{g2}.


\bigskip
\noindent {\bf Theorem 6.2}

{\it Suppose assumptions (A1)-(A3) (section 3) and (A4) (section 5) hold. Then, for $|g|<g_3$,  the time-periodic state $\omega_{g,s}^+, \; {\mathrm for} \; s\in [0,\tau),$ does not belong to $\N_\omega,$ ie, $\omega_{g,s}^+$ is not normal with respect to $\omega.$}

\bigskip

\noindent{\it Proof.}

First note that under the assumptions of this theorem, the results of Proposition 6.1 hold. In particular, $\tilde{K}_g$ has no real eigenvalues.

By construction,
\begin{equation}
\label{PerInv}
\omega_{g,s}^+\circ\a_g^{\pm\tau}=\omega_{g,s}^+.
\end{equation}

Since we are assuming that $\omega_{g,s}^+\in\N_\omega ,$
there exists a unique vector $\Omega_{g,s}^+$ in the natural positive cone
${\mathcal P}:=\{ aJa\Omega: a\in\M\}$ associated to the pair $(\M,\omega),$
such that
\begin{equation}
\label{Normal}
\omega_{g,s}^+(a)=\langle \Omega_{g,s}^+, a \Omega_{g,s}^+\rangle \; ,
\end{equation}
for $a\in\M.$\footnote{For a proof of this statement, see [BR],
Theorems 2.5.31 and 2.3.19.}
Now, \fer{StandEvol}, \fer{PerInv}, and \fer{Normal} imply that
\begin{equation}
\langle \overline{U}_g(\tau,0)\Omega_{g,s}^+, a \overline{U}_g(\tau,0)\Omega_{g,s}^+\rangle =  \langle \Omega_{g,s}^+, a \Omega_{g,s}^+\rangle ,
\label{PerExp}
\end{equation}
for $a\in\M.$ Using the facts that
\begin{equation}
[J,\overline{U}_g(t,t')]=0 \; ,
\end{equation}
where $J$ is the modular conjugation and $\overline{U}_g$ is the propagator generated by the standard Liouvillean, $J\Omega_{g,s}^+=\Omega_{g,s}^+$ since $\Omega_{g,s}^+\in{\mathcal P},$ $J^*=J,$ and $J\M J=\M',$ it follows that \fer{PerExp} also holds for $a\in\M'.$ Furthermore, we know that\footnote{see Corollary 2.5.32 in [BR].}
$$\overline{U}_g{\mathcal P}\subset {\mathcal P},$$
and that the linear span of ${\mathcal P}$ is dense in $\H.$ Let $\sfP_{g,s}^+$ be the orthogonal projection onto $\Omega_{g,s}^+,$ then $\sfP_{g,s}^+\in\M\vee\M'.$ Moreover, since $\pi=\pi^\S\otimes\pi^{\R_1}\otimes\cdots\otimes\pi^{\R_n},$ in the Araki-Wyss representation, $\M$ is a factor (of type $III_1;$ see [ArWy]), ie, $\M\cap\M'=\{{\mathbf C}\unit\}.$\footnote{Using the isomorphism between $\F^{\R_1}(L^2)\otimes\cdots\otimes\F^{\R_n}(L^2)$ and $\F(L^2_{\R_1}\oplus\cdots\oplus L^2_{\R_n}),$ one can proceed to show that $\M$ is a factor as in the case of the Araki-Wyss representation for a single reservoir of free fermions.}
Suppose that
$$\overline{U}_g(\tau,0)\Omega_{g,s}^+ = c_1\Omega_{g,s}^+ + c_2\Psi ,$$
where $\Psi\in Ran(1-\sfP_{g,s}^+),$ and $c_{1,2}$ are complex numbers to be determined. Choosing $a=(1-\sfP_{g,s}^+)$ in \fer{PerExp} gives
$$|c_2|^2=0.$$
Together with the fact that
$$\langle \overline{U}_g(\tau,0)\Omega_{g,s}^+, \overline{U}_g(\tau,0)\Omega_{g,s}^+\rangle =  \langle \Omega_{g,s}^+, \Omega_{g,s}^+\rangle=1,$$
it follows that $|c_1|^2=1.$ This implies that there exists $\lambda\in{\mathbf R}$ such that
\begin{equation}
\overline{U}_g(\tau,0)\Omega_{g,s}^+=e^{-i\lambda \tau}\Omega_{g,s}^+.
\end{equation}
For each fixed $s\in [0,\tau]$, we define
\begin{equation}
\phi_s(t):=e^{i\lambda t}\overline{U}_g(t,0)\Omega_{g,s}^+,
\end{equation}
then $\phi_s$ is an eigenfunction of $\tilde{K}_g$ with eigenvalue $\lambda.$ (This can be checked by looking at $\tilde{K}_g\phi_s.$)  However, this is in {\it contradiction} with the result of Proposition 6.1, and hence $\omega_{g,s}^+\not\in\N_\omega.$ $\Box$
\bigskip

We have the following result regarding the strict positivity of entropy production per cycle.

\bigskip
\noindent{\bf Theorem 6.3 (Positivity of entropy production)}

{\it Suppose assumptions (A1)-(A4) hold. Then the entropy production per cycle, after the state of the system has converged to a time-periodic state, is strictly positive, ie,
\begin{equation}
\Delta Ent := \int_0^\tau dt \omega^+_{g,t}(\delta_\omega (gV(t)))>0.
\label{entropypercycle}
\end{equation}
}
\bigskip

\noindent{\it Proof.} It follows from assumptions (A1)-(A4) and Theorem 5.1, that
$$\sup_{T\in {\mathbf R}^+}|\int_0^T dt \{\omega_{g,t\; mod \; \tau}^+ (\delta_\omega (gV(t)))-\omega\circ\a_g^t (\delta_\omega (gV(t)))  \}|<\infty.$$ Together with the result of Theorem 6.2, this implies that $\omega_{g,s}^+$ satisfies the assumptions of Proposition 2.1, and the entropy production per cycle after the state of the coupled system has converged to a time-periodic state, is strictly positive. $\Box$
\bigskip

Regarding the explicit computation of entropy production per cycle, \fer{entropypercycle},
we would like to make the following remark. Since it follows from Theorem 4.3 that $\tilde{\Omega}_{g,(0)}$ is analytic in $g$ for $g<g_1/2,$ one can expand $\omega^+_{g,s}$ to any order in the coupling $g,$ and compute an explicit expression for $\Delta Ent$ given in \fer{entropypercycle} up to this order in the coupling constant; (see also [MMS1,2] for a discussion of a perturbative approach for calculating entropy production in nonequilibrium steady-states).


\pagebreak
\bigskip
\noindent{\bf Appendix A}
\bigskip

\noindent{\bf Glued Hilbert space representation}
\bigskip

We want to show that
\begin{equation*}
\F(L^2({\mathbf R}^+;\B))\otimes \F(L^2({\mathbf R}^+;\B)) \cong \F(L^2 ({\mathbf R};\B)) \; .
\end{equation*}
Let $\Omega$ be the vacuum state in the fermionic Fock space $\F(L^2({\mathbf R}^+;\B)).$ For fermionic creation/annihilation operators on $\F(L^2({\mathbf R}^+; \B)),$
$$b^\# (f):= \int m(u)dud\sigma f(u,\sigma)b^\# (u,\sigma) \; , f \in L^2({\mathbf R}^+; \B),$$
define the creation/annihilation operators on $\F(L^2({\mathbf R}^+;\B))\otimes \F(L^2({\mathbf R}^+;\B))$ as
\begin{align*}
b_l^\# (f) := b^\# (f)\otimes \unit \; ; \\
b_r^\# (f) := (-1)^{N} \otimes b^\# (\overline{f}) \; ,
\end{align*}
where $\overline{\cdot}$ corresponds to complex conjugation. Note that $b_l$ and $b_r$ anti-commute. Let $\tilde{a}$ and $\tilde{a}^*$ be the annihilation and creation operators on the fermionic Fock space $\F(L^2({\mathbf R}^+;\B)\oplus L^2({\mathbf R}^+;\B)),$ such that they satisfy the usual CAR, and let $\tilde{\Omega}$ be the vacuum state in $\F(L^2({\mathbf R}^+;\B)\oplus L^2({\mathbf R}^+;\B)).$
An isomorphism between $\F(L^2({\mathbf R}^+;\B))\otimes \F(L^2({\mathbf R}^+;\B))$
and $\F(L^2({\mathbf R}^+;\B)\oplus L^2({\mathbf R}^+;\B))$ follows by the identification
\begin{align*}
b_l^\# (f)&\cong \tilde{a}^\# ((f,0)) ,\\
b_r^\# (g) &\cong \tilde{a}^\# ((0,g)) \; ,\\
\Omega\otimes\Omega&\cong \tilde{\Omega}.
\end{align*}

Now we claim that $\F(L^2({\mathbf R}^+;\B)\oplus L^2({\mathbf R}^+;\B))$ is isomorphic to
$\F(L^2 ({\mathbf R},du;\B))$. For $\phi, \psi \in {\mathbf R}$, consider the mapping
\begin{equation*}
j_{\phi,\psi } : L^2({\mathbf R}^+;\B)\oplus L^2({\mathbf R}^+;\B) \ni (f,g)\rightarrow h \in L^2 ({\mathbf R}du;\B) \; ,
\end{equation*}
such that
\begin{equation*}
h(u,\sigma) :=
\begin{cases}
e^{i\phi} \sqrt{m(u)} f(u,\sigma) \; , u\ge 0 \\
e^{i\psi} \sqrt{m(|u|)} g(|u|,\sigma) \; , u < 0
\end{cases} \; .
\end{equation*}
This mapping is an isometry, since
\begin{align*}
\| h \|^2_{L^2 ({\mathbf R},du;\B)} &= \| (f,g) \|^2_{ L^2({\mathbf R}^+;\B) \oplus L^2({\mathbf R}^+;\B) } \\
&= \int_{{\mathbf R}^+;\B} du d\sigma m(u) |f(u,\sigma)|^2 + \int_{{\mathbf R}^+;\B} du d\sigma m(u) |g(u,\sigma)|^2 \\
&= \| f \|^2_{L^2({\mathbf R}^+;\B)} + \| g \|^2_{L^2({\mathbf R}^+;\B)} \; .
\end{align*}
Moreover, the mapping $j_{\phi,\psi }$ is an isomorphism, since, for given
$h\in L^2 ({\mathbf R};\B)$, there exists a mapping
$j_{\phi, \psi}^{-1}: h\rightarrow (f,g)\in L^2({\mathbf R}^+;\B)\oplus L^2({\mathbf R}^+;\B)$, such that
\begin{align*}
f(u,\sigma)&:= \frac{e^{-i\phi}}{\sqrt{m(u)}}h(u,\sigma) , u>0 \; , \\
g(u,\sigma)&:= \frac{e^{-i\psi}}{\sqrt{m(|u|)}}h(|u|,\sigma) , u<0 \; .
\end{align*}


\begin{thebibliography}{ABCD}

\bibitem[Ar]{}Araki, H. : Relative Hamiltonian for faithful normal states of a von Neumann algebra, Publ. Res. Inst. Math. Sci. Kyoto Univ. 9, 165 (1973)

\bibitem[ArWy]{} Araki, H. and Wyss, W.: Representations of canonical anticommutation relations, Helv. Phys. Acta {\bf 37}, 136 (1964)

\bibitem[A-S]{} Abou Salem, W.: Nonequilibrium quantum statistical mechanics and thermodynamics, ETH-Diss. 16187 (2005)

\bibitem[A-SF1]{} Abou Salem, W. and Fr\"ohlich, J.: Status of the fundamental laws of thermodynamics, in preparation

\bibitem[A-SF2]{} Abou Salem,W. and Fr\"ohlich, J.: Adiabatic theorems and reversible isothermal processes, Lett. Math. Phys. {\bf 72}, 153-163 (2005)

\bibitem[BFS]{} Bach, V., Fr\"ohlich, J. and Sigal, I.M.: Return to Equilibrium, J. Math. Phys. {\bf 41 no 6}, 3985-4061 (2000)

\bibitem[BR]{} Bratteli, O. and Robinson, D.: Operator Algebras and Quantum Statistical Mechanics 1,2, Texts and Monographs in Physics, Springer-Verlag Berlin, 1987

\bibitem[D]{} Donald, M.J. : Relative Hamiltonians which are not bounded from above, J. Func. Anal. {\bf 91}, 143 (1990)

\bibitem[DJP]{} Derezi\'nski, J., Jaksi\'c, V. and Pillet, C.-A.: Perturbation theory of $W^*$-dynamics, Liouvilleans, and KMS-states, Rev. Math. Phys. {\bf 15}, 447-489 (2003)

\bibitem[FMSUe]{}Fr\"ohlich, J., Merkli, M.,  Schwarz, S., and Ueltschi, D.: Statistical mechanics of thermodynamic processes, in {\it A garden of quanta}, 345-363, World Sci. Publishing, River Edge, New Jersey, 2003

\bibitem[Ho]{} Howland, J.S.: Stationary scattering theory for time dependent Hamiltonians, Math. Ann. {\bf 207}, 315-335 (1974)

\bibitem[HP]{}Hunziker, W. and Pillet, C.-A.: Degenerate asymptotic perturbation theory, Commun. Math. Phys. {\bf 90}, 219 (1983)

\bibitem[Hu]{} Hunziker, W.: Notes on asymptotic perturbation theory for Schr\"odinger eigenvalue problems, Helv. Phys. Acta {\bf 61}, 257-304 (1988)

\bibitem[JP1]{} Jaksi\'c, V. and Pillet, C.A.: On a Model for Quantum Friction II. Fermi's Golden Rule and Dynamics at Positive Temperature, Commun. Math. Phys. {\bf 176}, 619-644 (1996)

\bibitem[JP2]{} Jaksi\'c, V. and Pillet, C.A.: On a Model for Quantum Friction III. Ergodic Properties of the Spin-Boson System, Commun. Math. Phys.  {\bf 178}, 627-651 (1996)

\bibitem[JP3]{} Jaksi\'c, V. and Pillet, C.-A.: Non-equilibrium steady states of finite quantum systems coupled to thermal reservoirs, Commun. Math. Phys. {\bf 226}, 131-162 (2002)

\bibitem[JP4]{}Jaksi\'c, V. and Pillet, C.-A.: A note on the
entropy production formula, Advances in differential equations and
mathematical physics, 175-180, Contemp. Math. {\bf 327}, American
Mathematical Society, Providence, RI, 2003

\bibitem[Ka1]{} Kato, T.: Perturbation theory for linear operators, Berlin: Springer, 1980

\bibitem[Ka2]{} Kato, T.: Linear evolution equations of hyperbolic
type, I.J. Fac. Sci. Univ. Tokyo Sect. IA {\bf 17}, 241-258 (1970)

\bibitem[MMS1]{} Merkli, M., M\"uck, M. and Sigal, I.M.: Instability of equilibrium states for coupled heat reservoirs at different temperatures, [axiv:math-ph/0508005]

\bibitem[MMS2]{} Merkli, M., M\"uck, M. and Sigal, I.M.: Theory of nonequilibrium stationary states as a theory of resonances. Existence and properties of NESS, [arxiv:math-ph/0603006]


\bibitem[PW]{} Pusz, W. and Woronowicz, S.L.: Passive states and KMS states for general quantum systems, Commun. Math. Phys. {\bf 58}, 273-290 (1978)

\bibitem[RS1,2]{} Reed, M. and Simon, B.: Methods of Modern Mathematical Physics, Vol. I (Functional Analysis), Vol. II (Fourier Analysis, Self-Adjointness), Academic Press, New York 1975

\bibitem[Rud]{} Rudin, W.: Real and Complex Analysis, 3rd
ed., Mc-Graw-Hill, New York, 1987


\bibitem[Ya1]{} Yajima, K.: Scattering theory for Schr\"odinger equations with potentials periodic in time, J. Math. Soc. Japan {\bf 29}, 729-743 (1977)

\bibitem[Ya2]{} Yajima, K.: Resonances for AC-Stark effect, Commun. Math. Phys. {\bf 87}, 331-352 (1982)

\bibitem[Yo]{} Yosida, K.: Functional Analysis, 6th ed., Springer-Verlag, Berlin, 1998


\end{thebibliography}
\end{document}